\documentclass[twocolumn, prc, amssymb, superscriptaddress,aps,preprintnumbers,amsmath,floatfix]{revtex4}
\usepackage{multirow}
\usepackage{graphicx}
\usepackage{dcolumn}
\usepackage{bm}
\usepackage{color}
\usepackage{amsmath}
\usepackage{wasysym}
\usepackage{float}
\usepackage{array,url,kantlipsum}
\usepackage{verbatim}
\usepackage{xcolor}
\usepackage{siunitx}
\usepackage{CJK}
\usepackage{dsfont}
\usepackage{textcomp}
\newcolumntype{P}[1]{>{\centering\arraybackslash}p{#1}}
\usepackage{array}
\usepackage{braket}
\usepackage{lipsum}
\usepackage{upgreek}
\usepackage{physics}
\usepackage[caption=false]{subfig}

\newcommand{\lrb}[1]{\left(#1 \right)} 
\newcommand{\lrsb}[1]{\left [#1 \right]} 
\newcommand{\lrcb}[1]{\left \{#1 \right \}}

\newcommand{\bfit}[1]{\bm{\mathit{#1}}}
\begin{document}


\title{Nuclear forward scattering of Bessel beams in $^{229}$Th:CaF$_2$}


\author{Alexander \surname{Franz}}
\email{alexander.franz@uni-wuerzburg.de}
\affiliation{University of W\"urzburg, Institute of Theoretical Physics and Astrophysics,  Am Hubland, 97074 W\"urzburg, Germany}

\author{Tobias Kirschbaum}
\affiliation{University of W\"urzburg, Institute of Theoretical Physics and Astrophysics,  Am Hubland, 97074 W\"urzburg, Germany}

\author{Adriana P\'alffy}
\email{adriana.palffy-buss@uni-wuerzburg.de}
\affiliation{University of W\"urzburg, Institute of Theoretical Physics and Astrophysics,  Am Hubland, 97074 W\"urzburg, Germany}


\date{\today}

\begin{abstract}
The coherent pulse propagation of a Bessel beam resonant to the 8.4 eV nuclear clock transition in $^{229}$Th-doped crystals is investigated theoretically. Due to the magnetic dipole character of the clock transition, Bessel beams which present non-uniform transverse profiles and carry orbital angular momentum might enhance excitation channels or offer new control degrees of freedom compared to standard plane waves.  
We model the nuclear forward scattering of a resonant Bessel beam pulse propagating through the crystal, extending an formalism based on the iterative wave equation for plane waves. Thereby we take into account the nuclear quadrupole splitting in the crystal, considering the possibility of multiple quantization axes and present results for scenarios involving a single nuclear transition and multiple simultaneously driven transitions, analyzing  temporal and spatial intensity patterns.  Our findings show that the propagation of Bessel beams can be used  to determine the relative distribution of different directions of quantization axes inside the crystal.

\end{abstract}

\maketitle

\section{Introduction}
\label{intro}
A particularly promising candidate for a nuclear clock is the first excited (isomeric) state of the isotope $^{229}$Th \cite{peik2003nuclear, peik2021nuclear, helmer1994excited, matinyan1998lasers,tkalya2003properties}. The corresponding nuclear transition lies at an energy of approximately $\sim 8.4\text{ eV}$ \cite{Seiferle2019, zhang2024dawn} in the vacuum-ultraviolet (VUV) regime, making it accessible to modern lasery systems. At the same time, the transition exhibits a long lifetime on the order of $\sim10^3 \text{ s}$, resulting in an exceptionally narrow linewidth. These properties make the $^{229}$Th isomer transition a promising candidate for secondary, independent frequency standard based on nuclear degrees of freedom. \newline
Two main approaches are currently pursued towards the realization of a $^{229}$Th nuclear clock \cite{peik2015nuclear}. The first approach is the single-ion nuclear clock \cite{peik2003nuclear, campbell2012single}. There, a thorium ion is confined in an ion trap. This offers precise control over the nucleus and its environment, but the experimental effort is technically demanding, requiring complex experimental techniques for ion trapping and cooling. The second approach is the solid-state nuclear clock, where $^{229}$Th atoms are doped into a VUV-transparent host crystal \cite{elwell2024laser, hudson2010, kazakov2012performance, jeet2015results, thirolf2024thorium, dessovic2014229thorium}. This approach allows one to address a macroscopic number of nuclei, even at room temperature, providing logistical benefits and a larger number of excited nuclei than in the trapped-ion approach. However, the crystal environment introduces systematic shifts, broadening mechanisms, and level splittings that must be understood and controlled.\newline
Only recently, the direct laser excitation of the $^{229}$Th isomer was accomplished in crystals after years of difficulty \cite{PRL2024, elwell2024laser, zhang2024dawn}. The long search was mainly caused by previously uncertain transition energy and the extremely narrow linewidth. In addition, the transition is not an allowed electric dipole ($\mathcal{E}1$) but has mixed magnetic dipole ($\mathcal{M}1$) and electric quadrupole ($\mathcal{E}2$) character. Standard plane waves couple comparatively weakly to electric dipole-forbidden transitions, which motivates exploring  whether structured light might enhance excitation channels or offer new control degrees of freedom. One such class of structured fields is vortex (or twisted) light. \newline
Vortex beams carry orbital angular momentum in addition to spin angular momentum along their direction of propagation \cite{allenPaper}. Compared to plane waves, they feature helical wavefronts and spatially inhomogeneous intensity profiles \cite{matula2013atomic, peshkov2017photoexcitation, yao2011orbital}. These properties have driven extensive research in the past decade, especially in atomic physics: vortex beams have been used to access higher-order multipole transitions beyond electric dipole \cite{afanasev2016high, afanasev2018atomic}, to generate spatially structured excitation patterns in atomic ensembles \cite{afanasev2013off, tw_hydrogenlike, sur_many, peshkov2018rayleigh, schulz2019modification, schmidt2023atomic}, and to investigate the propagation of resonant vortex light through ensembles of multi-level systems. \cite{hamedi2019transfer, mahdavi2020manipulation, meng2023coherent,hamedi2021ferris}. \newline
This success in atomic physics naturally motivates extending the idea to the photoexcitation of nuclei with vortex beams, in particular to the $^{229}$Th isomer transition. Previous theoretical studies have examined both the single-ion and solid-state nuclear clock scenarios involving twisted light \cite{kirschbaum2024photoexcitation}. For the single-ion case, important insights were gained into the mixed $\mathcal{M}1$+$\mathcal{E}2$ coupling and general aspects of nucleus-vortex beam interaction, such as spatial selection rules. However, no enhancement of the excitation probability compared to plane waves was identified. Similar conclusions were reached for the solid-state case.

An important aspect that has not yet been explored is the coherent propagation of resonant vortex beam pulses through a nuclear ensemble. In a crystal host, the nuclei form a resonant medium, making collective scattering effects relevant. In atomic physics, analogous situations have been studied for vortex beams propagating through multi-level systems \cite{hamedi2019transfer, mahdavi2020manipulation, meng2023coherent,hamedi2021ferris}. Due to the spatial structure of a vortex beam such propagation dynamics can potentially encode information about the internal crystal structure of $^{229}$Th-doped systems, such as the relative distribution of different doping orientations.
\newline
Existing theoretical studies of vortex beam propagation are typically based on steady-state approaches that neglect the full time dependence of the scattering process \cite{hamedi2019transfer, mahdavi2020manipulation, meng2023coherent,hamedi2021ferris}. As a result, time resolved phenomena such as dynamical or quantum beats cannot be described. They also do not provide a straightforward way to account for arbitrary orientations between the nuclear quantization axis and the light propagation direction. This aspect is particularly important in $^{229}$Th-doped crystals, where electric field gradients with varying orientation occur throughout the lattice and induce a quadrupole splitting of the clock transition \cite{dessovic2014229thorium}. \newline
For these reasons, this work develops a propagation formalism for vortex beams interacting with an ensemble of nuclei. We focus on a specific and widely used class of structured beams, namely Bessel beams, and extend the iterative wave equation (IWE) approach originally formulated for plane waves propagating in samples of M\"ossbauer nuclei \cite{shvyd1993coherent,shvyd1994perturbed, shvyd1999coherent}. The formalism is applied to the system $^{229}$Th:CaF$_2$ \cite{dessovic2014229thorium, beeks2023growth} to investigate the coherent propagation of resonant vortex pulses in a solid state nuclear medium. 

We first analyze the propagation dynamics for a single two level transition and different orientations of the quantization axis relative to the beam propagation direction. For parallel alignment, the transmitted field develops an inhomogeneous transverse intensity profile that reflects the spatial dependence of the Bessel beam interaction and remains constant in time. In contrast, for orthogonal alignment, the transverse profile exhibits temporal fluctuations driven by spatially varying time spectra. When all possible quantization axis orientations are included simultaneously, these fluctuations persist and, under suitable conditions, can be used to deduce the relative population of differently oriented nuclear sub ensembles. We investigate this scenario also in the non paraxial regime and find that scattering in the orthogonal alignment generates additional vortices.
We then extend the analysis to broadband excitation of the full hyperfine level scheme, where quantum beats dominate the temporal dynamics.

This paper is structured as follows. In Section~\ref{theory} we introduce the theoretical framework required to describe nuclear forward scattering of twisted light, including a brief overview of Bessel beams, the system of interest $^{229}$Th:CaF$_2$ and the plane wave IWE formalism. In Section~\ref{application} we generalize the propagation theory to Bessel beams and investigate the resulting dynamics for different orientations of the quantization axis. The analysis is first performed for a single two level transition and subsequently extended to the case of a multi level excitation. The paper concludes with a brief discussion in Section~\ref{sec:fin}.

\section{Theoretical Background} \label{theory}
This Section introduces the theoretical description of the Bessel beams, which are an analytically convenient form of twisted light. 
After giving a brief overview of the system of interest $^{229}$Th:CaF$_2$, we proceed by introducing the iterative wave equation formalism of the simple case of plane waves.

\subsection{Bessel beams }
\begin{figure}[h]
    \centering
    \includegraphics[width=0.8\linewidth]{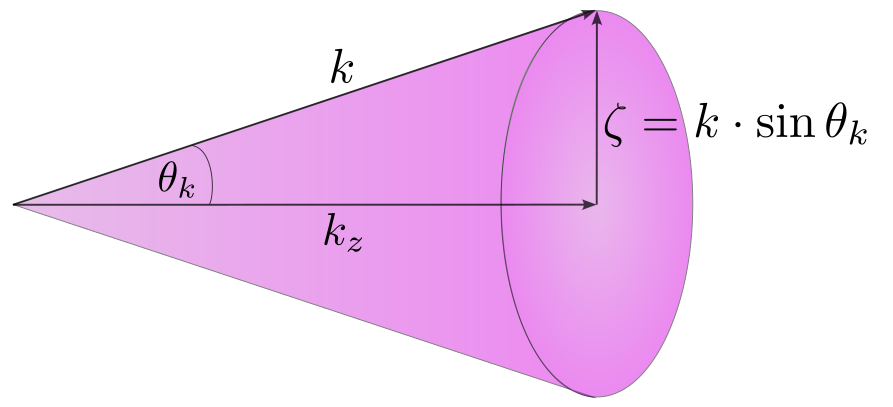}
    \caption{Bessel beam in momentum space as a superposition of plane waves spanning the surface of a cone with opening angle $\theta_k= \arctan \lrb{\zeta/k_z}$.}
    \label{fig:placeholder}
\end{figure}

We consider monochromatic Bessel beams propagating along the $z$-axis. In contrast to plane waves, Bessel beams carry a well defined projection of the total angular momentum (TAM) onto the propagation direction. This TAM generally contains contributions from spin angular momentum (SAM), associated with polarization, and orbital angular momentum (OAM), originating from the azimuthal phase structure of the field. Throughout this Section we assume free space propagation and work in the Coulomb gauge. 

A Bessel beam with transverse momentum $\zeta$, wave number $k$, helicity $\Lambda=\pm1$, and TAM projection $m_\gamma$ can be written as a coherent superposition of plane waves \cite{schulz2020generalized, peshkov_HG}
\begin{align}
    \bfit{A}_{\zeta m_\gamma}\lrb{\bfit{r},t}&=\bfit{A}_{\zeta m_\gamma}\lrb{\bfit{r}} e^{-i \omega t}
    \\ &=A_0 e^{-i \omega t}\int e^{i\bfit{k r}} \bfit{e}_{\bfit{k}\Lambda} a_{\zeta m_\gamma}\lrb{\bfit{k_\perp}} k_\perp \frac{dk_\perp}{2\pi} \frac{d\alpha_k}{2\pi} \ , \label{besselbeam}
\end{align}
where $\bfit{k}$ is $\lrb{\bfit{k}_\perp,k_z}$, $\alpha_k$ is the azimuthal angle of $\bfit{k}_\perp$, $e_{\bfit{k}\Lambda}$ denotes the circular polarization vector with helicity $\Lambda$ and $A_0$ is the intensity of the beam. The momentum distribution reads
\begin{align}
    a_{\zeta m_\gamma}\lrb{\bfit{k_\perp}} = (-i)^{m_\gamma} e^{i m_\gamma \alpha_k} \frac{2\pi}{\zeta} \delta\lrb{k_\perp-\zeta} \ ,
\end{align}
The delta distribution enforces $k_\perp=\zeta$, meaning that all contributing plane wave components lie on a cone with opening angle $\theta_k$ defined by 
\begin{align}
    \zeta= k \sin \theta_k
\end{align}
Thus, a Bessel beam is a coherent superposition of circularly polarized plane waves whose wave vectors span the surface of a cone. The azimuthal phase factor $e^{i m_\gamma \alpha_k}$ ensures a well defined TAM projection $m_\gamma$ along the $z$-axis. 

Since Eq.\eqref{besselbeam} is formulated in Coulomb gauge, the corresponding electric and magnetic fields are obtained as 
\begin{align}
    \bfit{E}_{\zeta m_\gamma}\lrb{\bfit{r},t}&=-i\omega A_0 e^{-i \omega t} \notag \\ &\cross \int e^{i\bfit{k r}} \bfit{e}_{\bfit{k}\Lambda} a_{\zeta m_\gamma}\lrb{\bfit{k_\perp}} k_\perp \frac{dk_\perp}{2\pi} \frac{d\alpha_k}{2\pi} \ , \label{Ebessel} \\
    \bfit{B}_{\zeta m_\gamma}\lrb{\bfit{r},t}&= i A_0 e^{-i \omega t} \notag \\ &\cross \int e^{i\bfit{k r}} \lrb{\bfit{k}\cross \bfit{e}_{\bfit{k}\Lambda}} a_{\zeta m_\gamma}\lrb{\bfit{k_\perp}} k_\perp\frac{dk_\perp}{2\pi} \frac{d\alpha_k}{2\pi} \ . \label{Bbessel}
\end{align}
The time-averaged Poynting vector is given by
\begin{align}
    \bfit{S}\lrb{\bfit{r}}=\frac{1}{2 \mu_0}\Re\lrb{\bfit{E}\lrb{\bfit{r}}\cross \bfit{B^*}\lrb{\bfit{r}}} \ . \label{poyntingE} 
\end{align}
Unlike plane waves, the longitudinal wave component $S_z$ depends on the transverse radius $r_\perp$, resulting in a structured intensity profile \cite{surzhykov2016intensity}.

To obtain the spatial structure explicitly, we expand the polarization vector in a spherical basis according to
\begin{align}
    \bfit{e}_{\bfit{k}\Lambda}=\sum_{m_s=-1}^1 c_{m_s}e^{i \alpha_k (m_\gamma - m_s)} \bfit{e}_{m_s} \ .
\end{align}
The coefficients depend on the cone angle $\theta_k$
\begin{align}
    c_0&=-\frac{1}{\sqrt{2}}\sin\theta_k,  \\  
    c_{\pm1}&=\frac{\pm \Lambda}{2} \lrb{1 \pm \Lambda \cos \theta_k}.
\end{align}
Using the integral representation of the Bessel function \cite{abramowitz1965handbook}
\begin{align}
    \int_0^{2\pi} e^{i l a} e^{i \pm x \cos\lrb{b-a}} da = 2\pi (\pm i)^l e^{ilb} J_{l}\lrb{x} \ . \label{besselrep}
\end{align}
the vector potential becomes 
\begin{align}
    \bfit{A}_{\zeta m_\gamma}\lrb{\bfit{r}} &= \frac{A_0}{2\pi} (-i)^{m_\gamma} e^{i k_z z} \notag \\
    &\cross \sum_{m_s=-1}^{m_s=1} \int e^{i \zeta r_\perp \cos \lrb{ \alpha_k-\alpha_r}} e^{i(m_\gamma -  m_s) \alpha_k } \bfit{e}_{m_s} d\alpha_k \notag \\
    &= A_0 e^{i k_z z} 
    \sum_{m_s=-1}^{m_s=1} (-i)^{m_s} c_{m_s} \notag \\ 
    &\cross J_{m_\gamma-m_s}\lrb{\zeta r_\perp} e^{i (m_\gamma - m_s) \alpha_r} \bfit{e}_{m_s} \ . \label{realspaceBB}
\end{align}
This expression explicitly exhibits the radial dependence through Bessel functions and the azimuthal phase structure. Helicity, TAM and opening angle can modify spatial excitation profiles in light-matter interaction \cite{kirschbaum2024photoexcitation}, and therefore directly affect the coherent propagation dynamics discussed in this work.
While most experimental realizations of twisted light operate in the paraxial regime, i.e. small opening angles, the non-paraxial regime provides additional degrees of freedom that can significantly affect the interaction with matter. In the following, we primarily focus on the paraxial regime, while selected results for larger opening angles are discussed to highlight qualitative differences in the propagation dynamics.

\subsection{The Th:CaF$_2$ crystal}

\begin{figure*}
    \centering
    \subfloat[\label{fig:zconfigsingletrans}]{\includegraphics[width=\linewidth]{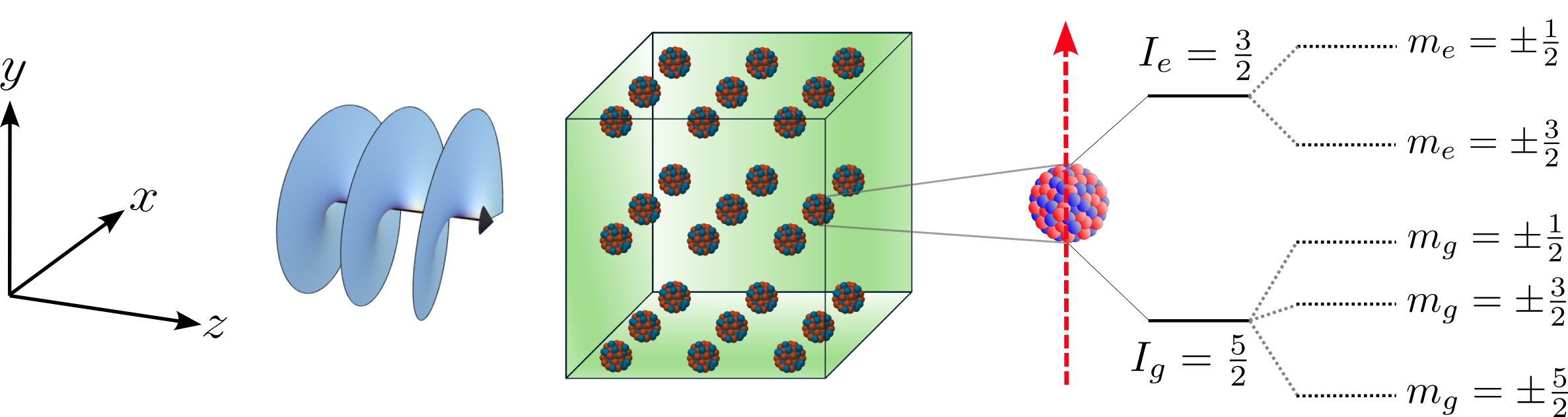}}
    \vspace{1em}
    \subfloat[\label{fig:zconfigsingletrans}]{\includegraphics[width=\linewidth]{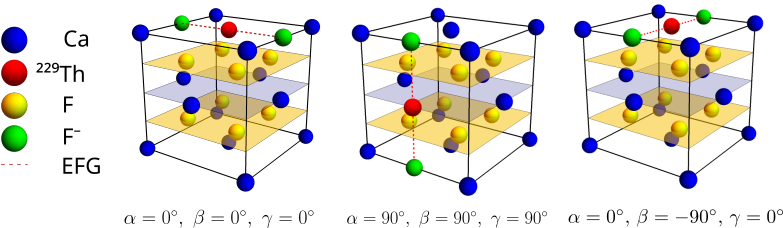}} 
    \caption{(a) A Bessel beam is propagating along the $z$-axis through a $^{229}$Th:CaF$_2$ sample. An electric field gradient inside the crystal leads to quadrupole splitting, giving rise to a many-level system. (b) The depicted  configuration corresponds to the 180$^\circ$-doping geometry. The three possible unit cell orientations are illustrated. The corresponding Euler angles with respect to the laboratory frame in the upper row are indicated below each orientation. }
    \label{fig:NFS_Bessel}
\end{figure*}
The  $^{229}$Th:CaF$_2$  crystal is an interesting candidate for a solid state nuclear clock \cite{kazakov2012performance, zhang2024dawn,dessovic2014229thorium}.
The host crystal CaF$_2$ offers an ideal environment for doping $^{229}$Th due to its band gap of $\approx11 \ \text{eV}-12 \ \text{eV}$ \cite{rubloff1972far, barth1990dielectric, tsujibayashi2002spectral}. This makes it transparent at the wavelength of the clock transition, preventing light interactions with the host crystal. Furthermore, large doping densities up to $N\approx10^{18} \frac{1}{\text{cm}^3}$ can be achieved \cite{beeks2023growth}. Thus, it is possible to interrogate more nuclei at the same time to obtain an improved clock stability.

Due to an electric field gradient (EFG) in this environment, the thorium nucleus experiences an quadrupole splitting described by the Hamiltonian \cite{collins1967electric}
\begin{align}
    \hat{H}_{E2}=\frac{eQV_{zz}}{4I(2I-1)}\left[3\hat{I}_z^2-\hat{I}^2+\frac{\eta}{2}\left( I_+^2 + I_-^2 \right) \right],
\end{align}
where $Q_g=3.11$ b \cite{campbell2012single,bemis1988coulomb} and $Q_e=1.8$ b \cite{tkalya_laser, thielking2018laser} are the quadrupole moments of the ground and isomer state, respectively, with $\text{b}=10^{-24}\text{ cm}^2$, and $e$ the elementary charge. $V_{zz}$ is the dominant component of the electric field gradient at the thorium nuclei, while $I_e$ and  $I_g$ once more denote the nuclear spin angular momenta of the excited and ground states, respectively. The operators $\hat{I}$ and $\hat{I_z}$ represent the total angular momentum operator and its projection on the quantization axis, respectively, and $\hat{I}_\pm$  are the corresponding raising and lowering operators. The asymmetry parameter $\eta=\frac{V_{xx}-V_{yy}}{V_{zz}}$ describes the deviation from axial symmetry in the EFG, where $V_{xx}$ and $V_{yy}$ are the smaller components of the field gradient.

The origin of the EFG lies in the difference in valence electrons between Ca and Th. During the doping process, one Ca atom in the unit cell of CaF$_2$ is substituted with a $^{229}$Th atom, as illustrated in Fig.~\ref{fig:NFS_Bessel}. Since Ca and Th differ in their number of valence electrons, compensating charge configurations form in order to maintain charge neutrality. Using density functional theory (DFT), one can determine the energetically favorable doping configurations. However, it is important to note that these theoretical configurations do not provide the full picture since recent experimental results reveal more energy levels than predicted, suggesting a mixture of multiple configurations \cite{zhang2024dawn}.\newline 
In this work, we consider the energetically favorable configurations predicted in \cite{dessovic2014229thorium}, the so-called 180$^\circ$-configuration, shown in Fig.~\ref{fig:NFS_Bessel}. Here, two additional fluorine ions next to the thorium atom cause an EFG with $V_{zz}=-296.7$V\r{A}. Due to the symmetric arrangement of the charges, the asymmetry parameter vanishes, i.e. $\eta=0$. Taking the direction of the EFG as the quantization axis leads to the level structure depicted in Fig.~\ref{fig:NFS_Bessel}. \newline 
Due to the cubic symmetry of the  CaF$_2$ crystal, not all unit cells containing thorium exhibit the same EFG orientation. Instead, there are three distinct, mutually orthogonal EFG directions, all of which must be taken into account when describing the propagation process (see Fig.~\ref{fig:NFS_Bessel}). In order to describe these orientations we employ Euler angles with respect to the laboratory frame using the convention of \cite{edmonds1996angular}.

\subsection{Theory of nuclear forward scattering for plane waves}
We consider the propagation of a weak resonant light pulse through a macroscopic ensemble of $^{229}$Th nuclei embedded in CaF$_2$. The nuclei are confined in the Lamb-Dicke regime, i.e., the recoil energy is much less than the energy required to create a phonon in the lattice \cite{dicke1953effect, Daslambdicke2012, hudson2010}. Excitation and emission therefore occur recoillessly, and the emitted photons remain resonant with the nuclear transition. This enables coherent multiple scattering and gives rise to nuclear forward scattering (NFS). 

Throughout this work, we assume the linear response regime, i.e., the excitation probability per nucleus remains small. In this limit, the dynamics can be derived from Maxwell's equations in matter together with the induced nuclear current density following the approach derived in \cite{shvyd1999coherent}.
Under the slowly varying envelope approximation (SVEA) and for propagation along the $z$-axis, the electirc field envelope $\bfit{E}\lrb{z,t}$ satisfies

    \begin{align}
\left(
\partial_z + \frac{1}{c}\partial_t
\right)
\bfit{E}(z,t)=
i \frac{\mu_0 \omega}{2}
\bfit{J}(z,t),
\end{align}
where $\bfit{J}\lrb{z,t}$ is the macroscopic nuclear current density induced by the field. Introducing dimensionless variables
\begin{align}
\tau =\Gamma t , \ \xi =\frac{N \sigma z}{4},
\end{align}
the time is scaled by the total radiative decay rate $\Gamma$, and the spatial coordinate expressed in terms of the effective thickness $\xi$. Here, $N$ is the nuclear number density, $\sigma$ is the total resonant absorption cross section ans $z$ the propagation distance. In these units, the electric field can be written as an infinite scattering series,
\begin{align}
   & \bfit{E}(\xi,\tau)=\sum_{p=0}^\infty \bfit{E}_p(\xi,\tau) = \sum_{p=0}^\infty \frac{(-\xi)^p}{p!} \bfit{E}_p(\tau)
\end{align}
with initial condition 
\begin{align}
    \bfit{E}_0(\tau)=\bfit{E}_{in}(\tau), 
\end{align}
where $\bfit{E}_{in}(\tau)$ denotes the incident pulse envelope. The recursive relation for the $p$-th scattering order reads
\begin{align}
    \bfit{E}_p(\tau) &= -\sum_l \bfit{j}_l\lrb{\bfit{k}} e^{-i\Omega_l\tau-\frac{\tau}{2}-\frac{\gamma_c}{\Gamma} \tau} \notag \\
   &\cross \int_{-\infty}^\tau d\tau' \bfit{j}^*_l\lrb{\bfit{k}} e^{i\Omega_l\tau+\frac{\tau'}{2}+\frac{\gamma_c}{\Gamma} \tau} \bfit{E}_{p-1}(\tau') .\label{eq:IWEplane}
\end{align}
This equation is known in the literature as the iterative wave equation (IWE).
The index $l$ labels the hyperfine transitions between ground and excited sublevels. The dimensionless frequency shift
\begin{align}
    \Omega_l=\frac{\omega_l-\omega_0}{\Gamma}
\end{align}
 accounts for quadrupole splitting relative to the unperturbed transition frequency $\omega_0$.
 The reduced nuclear current matrix elements for magnetic dipole transitions are given by
 \begin{align}
    \bfit{j_l\lrb{k}} &= \sqrt{3}  \sum_{q=0,\pm1} (-1)^q  
    \begin{pmatrix}
  I_g & 1 & I_e \\
  -M_g & q & M_e
\end{pmatrix} \left(\frac{\mathbf{k}}{|\mathbf{k}|} \times \mathbf{n}_{-q} \right) \label{eq:currentdensity}\\
\bfit{j_l^*\lrb{k}} &= \sqrt{3}  \sum_{q=0,\pm1} (-1)^q  
    \begin{pmatrix}
  I_g & 1 & I_e \\
  -M_g & q & M_e
\end{pmatrix} \left(\frac{\mathbf{k}}{|\mathbf{k}|} \times \mathbf{n}_{-q}^*\right)  \label{eq:currentdensitystar}
\end{align}
where $I_g$, $I_e$ are the nuclear spins of ground and excited states, $M_g$, $M_e$ denote their quantum numbers, $\bfit{n}_q$ with $q \in \lrcb{-1,0,1}$ form the spherical basis and the Wigner $3j$-symbol ensures angular momentum conservation and determines the selection rules \cite{edmonds1996angular}. The factor $e^{-\tau/2-\frac{\gamma_c}{\Gamma} \tau}$ in Eq.\eqref{eq:IWEplane} describes the natural radiative decay of the nuclear states and incorporates the decoherence effects inside the crystal (see. Appendix~\ref{app:decoherence}). 

Eq.~\eqref{eq:IWEplane} has an interpretation in terms of multiple coherent scattering: The $0$-th order is the incident field without interaction, the first order describes the single resonant scattering event and the second order is the re-excitation by the emitted field etc. Each successive term corresponds to a higher scattering order in the forward direction. 

For a single transition and a delta-like excitation pulse
\begin{align}
    \bfit{E}_{in}\lrb{\tau}=\bfit{E}_0 \delta \lrb{\tau},
\end{align}
the series can be evaluated analytically, yielding
\begin{align}
       \bfit{\bfit{E}}\lrb{\xi,\tau} &= \bfit{E}_0 \delta\lrb{\tau} \notag \\&- \xi \lrb{\bfit{E}_0 \cdot \bfit{j}_l(\bfit{k})} \frac{J_1\lrb{2 \sqrt{\xi |\bfit{j}_l(\bfit{k})|^2 \tau}}}{\sqrt{\xi |\bfit{j}_l(\bfit{k})|^2 \tau}} \bfit{j}_l(\bfit{k}) \notag \\
        &\cross e^{-i\Omega_l\tau-\frac{\tau}{2}-\frac{\gamma_c}{\Gamma} \tau}\ . \label{eq:IWEanalytical}
\end{align}
The Bessel function dependence on $\sqrt{\xi \tau}$ reflects the collective nature of nuclear forward scattering and the coherent build up of the scattered field with increasing effective thickness.

\section{Propagation dynamics} \label{application}
In this Section, we combine our knowledge about Bessel beams and the NFS of plane wave pulses in $^{229}$Th:CaF$_2$ to develop an IWE formalism to Bessel beam pulses. We then apply this formalism for two examples. First, we analyze a single transition and investigate how different orientations of the quantization axis modify the forward scattering dynamics. Second, we extend the treatment to the full hyperfine level scheme of $^{229}$Th:CaF$_2$, including all allowed transitions and EFG orientations. 

\subsection{IWE with Bessel beams}
We consider a Bessel beam propagating along the $z$-axis through the crystal. The pulse duration is assumed to be much shorter than the nuclear lifetime, such that the temporal envelope can be approximated by a delta distribution. The incident electric field is therefore written as
\begin{align}
    \bfit{E}^{BB}_{m_\gamma \Lambda,in}\lrb{\bfit{r},t} &= E_{in} \delta\lrb{t} \notag \\
    &\cross \int_0^{2\pi}  e^{i \lrb{\bfit{k r}-\omega t}} \bfit{e}_{\bfit{k}\Lambda} (-i)^{m_\gamma} e^{i m_\gamma \alpha_k} \frac{d \alpha_k}{2\pi} \ .
\end{align}
We exploit the momentum space representation of a Bessel beam and treat each plane wave component separately within the IWE formalism. We therefore express the propagated electric field as 
\begin{align}
   \bfit{E}^{BB}_{m_\gamma \Lambda}\lrb{\bfit{r},t} &= \int_0^{2\pi} \mathcal{E} \lrb{\bfit{k}, \bfit{r},t} e^{i \lrb{\bfit{k r}-\omega t}} \notag \\
   &\cross \bfit{e}_{\bfit{k}\Lambda} (-i)^{m_\gamma} e^{i m_\gamma \alpha_k} \frac{d \alpha_k}{2\pi} \ . \label{eq:Besselscattering}
\end{align}
where $\mathcal{E} \lrb{\bfit{k}, \bfit{r},t}$ denotes the individual envelope of a plane wave propagating along the cone surface. For each $\bfit{k}$, the envelope $\mathcal{E}$ satisfies the IWE for plane wave propagation. However, the effective thickness must be modified to account for the increased propagation path inside the medium. Since plane wave component propagates under an angle $\theta_k$ relative to the $z$-axis, the effective thickness transforms as 
\begin{align}
    \xi \rightarrow \frac{\xi}{\cos \theta_k} .
\end{align}
The magnetic field follows from Eq.~\eqref{Bbessel} as
\begin{align}
     \bfit{B}^{BB}_{m_\gamma \Lambda}\lrb{\bfit{r},t} &= \frac{1}{\omega} \int_0^{2\pi} \mathcal{E} \lrb{\bfit{k}, \bfit{r},t} e^{i \lrb{\bfit{k r}-\omega t}} \lrb{\bfit{k} \cross \bfit{e}_{\bfit{k}\Lambda}} \notag \\
     &\cross (-i)^{m_\gamma} e^{i m_\gamma \alpha_k} \frac{d \alpha_k}{2\pi} \ .
\end{align}
From the electric and magnetic fields, the Poynting vector is obtained using Eq.~\eqref{poyntingE}. Its $z$-component provides the spatially and temporally resolved transverse intensity.

\subsection{Single two-level transition}
We first apply the generalized formalism to a single nuclear transition. Specifically, we consider the transition from $\ket{\frac{5}{2}, \frac{5}{2}}$ to $\ket{\frac{3}{2}, \frac{3}{2}}$. The driving field is taken to be Bessel beam with helicity $\Lambda=-1$ and TAM $m_\gamma=1$. The beam propagates along the $z$-axis. We restrict our analysis mainly to the paraxial regime and choose the opening angle $\theta_k=5^\circ$. As crystal parameters, we use $N=10^{18}\frac{1}{\text{cm}^3}$, $\sigma \approx 2.4\cdot 10^{-10} \text{cm}^2$, $L=1 \ \text{cm}$ and $\Gamma\approx 10^{-3} \frac{1}{\text{s}}$, resulting in the effective thickness $\xi\approx6 \cdot 10^7$
As discussed previously, the EFG inside the crystal and therefore the quantization axis of the system does not have a unique direction. Therefore, the propagation problem must be solved separately for quantization axes aligned along the $x$- $y$- and $z$-axes.
\subsubsection{EFG parallel to beam propagation axis} \label{sec:singlepar}
\begin{figure}[h]
    \centering
    \subfloat[\label{fig:zconfigsingletrans}]{\includegraphics[width=\linewidth]{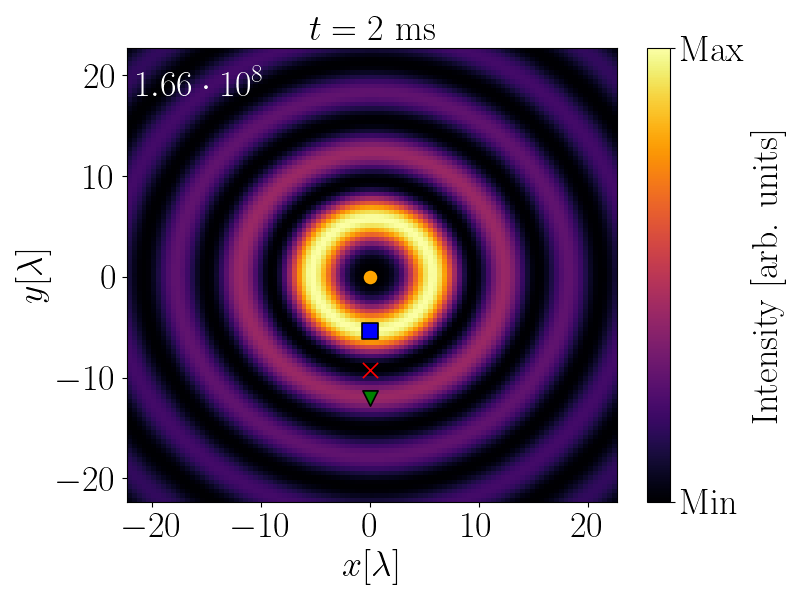}}
    \vspace{1em}
    \subfloat[\label{fig:zconfigsingletime}]{\includegraphics[width=\linewidth]{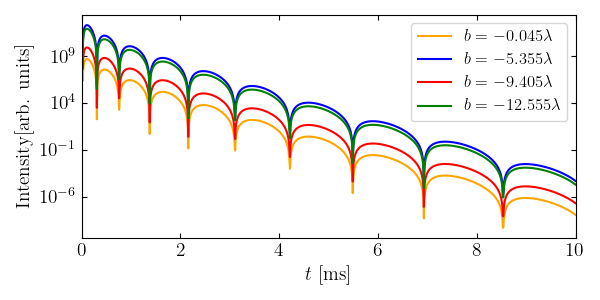}} 
    \caption{NFS result for a Bessel beam propagating parallel to the quantization axis ($\theta_k=5^\circ$). (a) shows the transverse intensity profile at the end of a $^{229}$Th:CaF$_2$ crystal at time $t=2 \ \text{ms}$, displaying an annular pattern. Maximum intensity is depicted in the upper left corner in arb. units. In (b) intensity time spectra are shown for the marked positions in the transverse plane (yellow circle: $b=-0.045 \lambda$, blue square: $b=-5.355 \lambda$, red cross: $b=-9.405 \lambda$, green triangle: $b=-12.555 \lambda$). All curves exhibit the same qualitative behaviour, sharing an identical dynamical beat frequency.}
    \label{fig:zconfigsingle}
\end{figure}
The results are depicted in Fig.~\ref{fig:zconfigsingle}. As expected, the transverse intensity profile shows an inhomogeneous profile due to the spatially selective interaction of the Bessel beam with a nucleus, see Appendix~\ref{app:interaction} for further information. This can be verified by evaluating time spectra at different points in the transverse plane, as illustrated in Fig.~\ref{fig:zconfigsingletime}. The qualitative behaviour and the beat frequencies are identical throughout, only the overall intensity changes. In fact, the dynamics is equivalent to that of a plane wave with changed effective thickness. This becomes evident by using Eq.~\eqref{eq:Besselscattering} to obtain the analytical solution
\begin{align}
   \bfit{E}^{BB}_{m_\gamma \Lambda}\lrb{\bfit{r},t} = &\int_0^{2\pi} \tilde{\xi} \lrb{\bfit{E}_0 \cdot \bfit{j}^*(\bfit{k})} \frac{J_1\lrb{2 \sqrt{\tilde{\xi} |\bfit{j}(\bfit{k})|^2 \tau}}}{\sqrt{\tilde{\xi} |\bfit{j}(\bfit{k})|^2 \tau}} \bfit{j}(\bfit{k}) \notag \\ &\cross  e^{i \lrb{\bfit{k r}-\omega t}} (-i)^{m_\gamma} e^{i m_\gamma \alpha_k} \frac{d \alpha_k}{2\pi} e^{-i\Omega_l\tau-\frac{\tau}{2}-\frac{\gamma_c}{\Gamma} \tau} \ , \label{eq:IWEanalyticalBB}
\end{align}
where $\tilde{\xi}=\frac{\xi}{\cos \lrb{\theta_k}}$a and we omit the zeroth scattering order since we consider times $t>0$. The oscillatory behaviour of the Bessel function $J_1$ defines the temporal beat structure, which is influenced by the modulus squared of the nuclear current density $|\bfit{j}(\bfit{k})|^2$. Using Eqs.~\eqref{eq:currentdensity} and \eqref{eq:currentdensitystar}, we obtain for the selected transition 
\begin{align}
    |\bfit{j}(\bfit{k})|^2 = \frac{1}{8}\lrb{3 + \cos \lrb{2 \theta_k}} \ .
\end{align}
This expression is independent of $\alpha_k$, meaning the integral over the cone does not modify the beat frequency. Consequently, the time spectrum is position independent across the transverse plane. This observation also holds for other individual transitions in the quadrupole-split level scheme, where the same angular independence is found (see Appendix~\ref{app:currents}).
While the time dependent dynamics are homogeneous across the beam profile, the spatial distribution of the scattered intensity shows a behaviour distinct from that of the input beam.
Regions of higher intensity correspond to larger excitation amplitudes and stronger interaction with the light field. This indicates that the spatial pattern of the scattered light should qualitatively follow the spatial dependence of the interaction matrix element for the Bessel beam. To confirm this, we compare the normalized radial scattered intensity with the normalized radial absorption profile of a nucleus-Bessel beam interaction (see Appendix~\ref{app:interaction}), shown in Fig.~\ref{fig:radmatrixcompzconfig}. 

\begin{figure}
    \centering
    \includegraphics[width = \columnwidth]{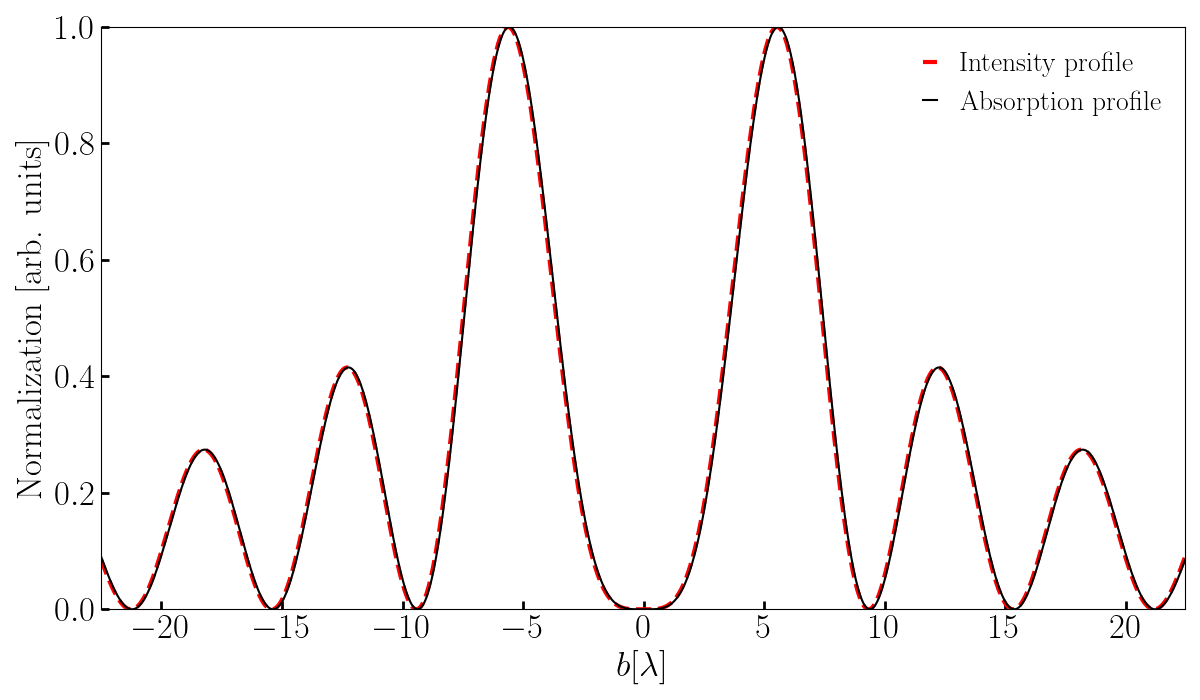}
    \caption{The normalized intensity profile is compared to the normalized absorption profile, showing perfect agreement. The intensity profile does not change in time on a qualitative level.}
    \label{fig:radmatrixcompzconfig}
\end{figure}

The curves match perfectly, demonstrating that in this case the transverse structure of the scattered field is entirely determined by the spatial dependence of the light-nucleus interaction. However, this depends on the type of transition. For a $\Delta m=0$-transition for example, the emission process modifies the spatial structure of the scattered field such that it does not overlap with the absorption profile, see Appendix~\ref{app:deltam0} for further discussion.

\subsubsection{EFG perpendicular to beam propagation axis}
\begin{figure}[h]
    \centering
    \subfloat[\label{fig:xconfigtransintensity}]{\includegraphics[width=\linewidth]{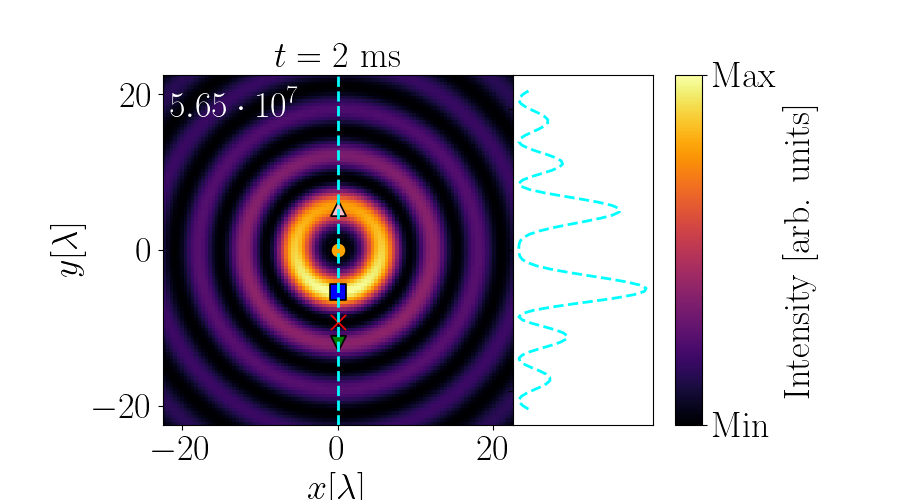}}
    \vspace{1em}
    \subfloat[\label{fig:xconfigtimespectra}]{\includegraphics[width=\linewidth]{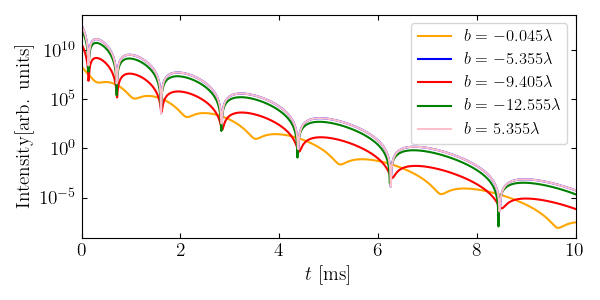}} 
    \caption{NFS result for a Bessel beam propagating orthogonal to the quantization axis ($\theta_k=5^\circ$). (a) shows the transverse intensity profile at the end of a $^{229}$Th:CaF$_2$ crystal at time $t=2 \ \text{ms}$, displaying an annular pattern. Maximum intensity is depicted in the upper left corner in arb. units. In (b) intensity time spectra are shown for the marked positions in the transverse plane (yellow circle: $b=-0.045 \lambda$, blue square: $b=-5.355 \lambda$, red cross: $b=-9.405 \lambda$, green triangle: $b=-12.555 \lambda$, pink triangle: $b=-5.355 \lambda$).}
    \label{fig:zwei-bilder}
\end{figure}
\begin{figure}
    \centering
    \includegraphics[width = \columnwidth]{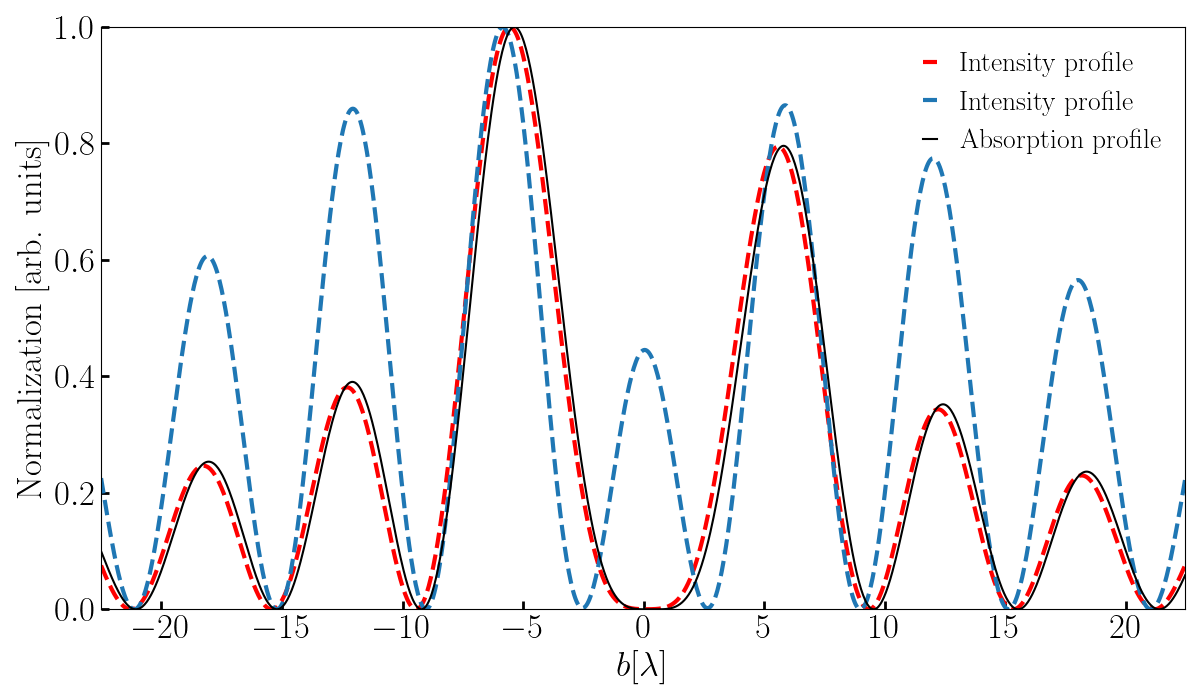}
    \caption{The normalized absorption profile (black curve) is compared to the normalized intensity profile at times $t=2$ ms (red curve) and $t=0.8$ ms (blue curve). For most times the intensity exhibits the profile of the red curve which coincides with the absorption profile. At certain times the agreement breaks down and one obtains deviations like the blue curve. This behaviour arises due to the spatial dependence of the beat frequency. The profiles are taken along the $y$-axis.}
    \label{fig:radmatrixcompxconfig}
\end{figure}

As discussed in the beginning of this Section, we must also consider the propagation for an EFG oriented along the $x$- and $y$-directions. We denote in the following these cases as "$x$-orthogonal orientation" and "$y$-orthogonal orientation", respectively. We proceed in analogy to the parallel orientation case. The only modification is that the set of vectors $\bfit{n}_q$, which appear in the calculation of $|\bfit{j}(\bfit{k})|^2$, must now be rotated using the Euler angles corresponding to the respective orientation, see Fig~\ref{fig:NFS_Bessel}. Since the qualitative results are similar for both $x$- and $y$-orthogonal orientations, we focus here on the $x$-axis orientation, defined by Euler angles $\alpha=0,\beta=-\frac{\pi}{2},\gamma=0$. 
In Fig.~\ref{fig:xconfigtransintensity}, the transverse intensity profile at time $t=2 \ \text{ms}$ is shown. As in the parallel orientation case, an annular pattern is observed. However, in this case a slight intensity gradient appears along the $y$-axis. It occurs due to the altered interaction profile in the $x$-orthogonal orientation. This is emphasized by a comparison along the $y$-axis of the scattered intensity and the absorption profile in Fig.~\ref{fig:radmatrixcompxconfig}. However, 
 the profile undergoes slight changes in time, which is indicated by showing time spectra at different positions in Fig~\ref{fig:xconfigtimespectra}. The curves at points with relatively low intensity differ from the ones with high intensity. To identify the origin of this spatial variation, we calculate the modulus squared of the nuclear current density 
\begin{align}
    |\bfit{j}(\bfit{k})|^2 = \frac{1}{8} \lrb{2 \cos^2 \lrb{\theta_k} + \lrb{3 + \cos \lrb{2 \alpha_k}} \sin^2\lrb{\theta_k} }\ . \label{eq:modulusnuccurrentx}
\end{align}
In contrast to the parallel case, this expression now depends on the azimuthal angle $\alpha_k$. Consequently, the Bessel function in Eq.~\ref{eq:IWEanalyticalBB} remains inside the integral, introducing a spatial dependence in the beat frequency. 

\subsubsection{Mixed orientations}
\begin{figure*}
    \centering
    \includegraphics[width=\linewidth]{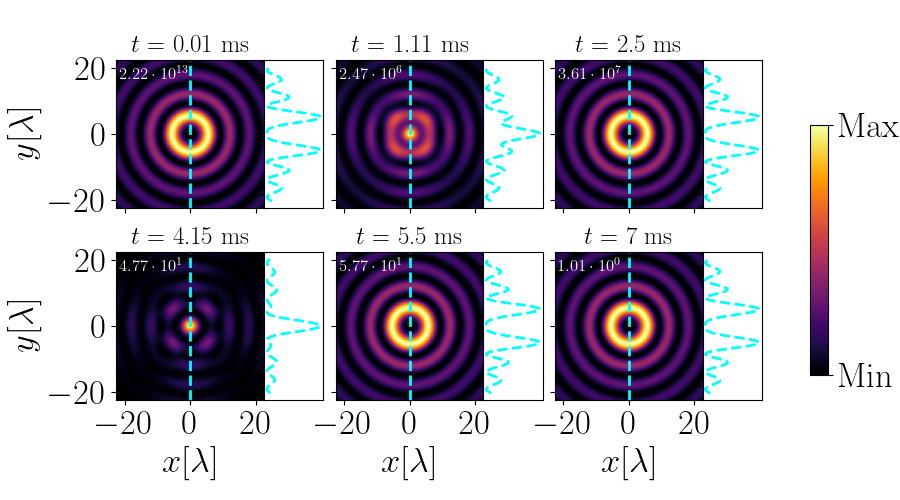}
    \caption{Transverse intensity profiles at different times for a Bessel beam propagating under a uniform distribution of quantization axis orientations ($\theta_k=5^\circ$). Temporal fluctuations in the intensity profile are observed, arising from different dynamical beat frequencies associated with individual orientations. Respective maximum intensities are depicted in the upper left corner in arb. units.}
    \label{fig:allconfigouniformosci}
\end{figure*}
\begin{figure*}
    \centering
    \includegraphics[width = \textwidth]{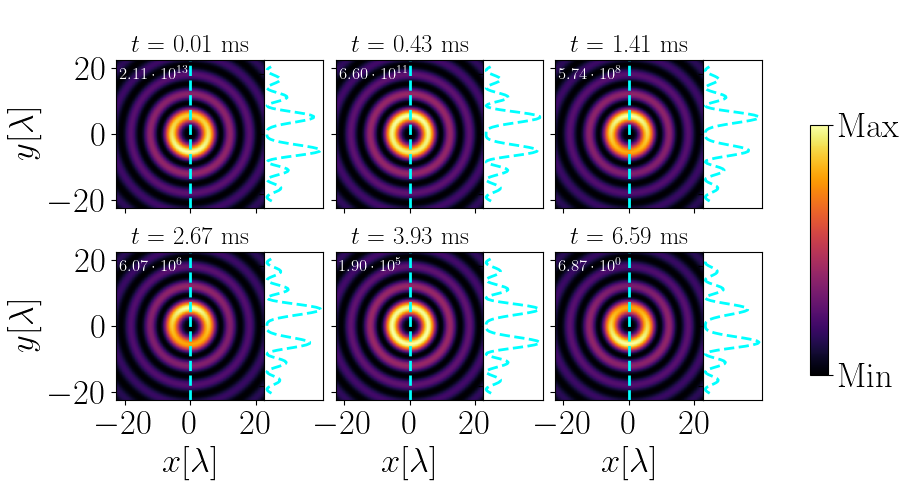}
    \caption{Transverse intensity profiles at different times for a Bessel beam propagating through a medium with electric field gradients pointing predominantly in $x$-direction ($\theta_k=5^\circ$). Temporal fluctuations in the intensity profile are observed, arising from the unequal distribution of parallel and anti-parralel aligned $x$-axis configurations. Respective maximum intensities are depicted in the upper left corner in arb. units.}
    \label{fig:allconfigxweightedosci}
\end{figure*}
To model the propagation in a realistic crystal environment, we simultaneously consider all orientations of the quantization axis inside the crystal. It is important to clarify that there are, in fact, six distinct orientations, as it makes a difference whether the EFG is aligned parallel or anti-parallel to a given axis. However, the anti-parallel configuration simply leads to the results of the parallel alignment but with a beam of opposite helicity. \newline
The total scattered electric field accounting for all doping orientations can be written as the superposition of the electric fields corresponding to each individual configuration
\begin{align}
    \bfit{E}^{BB}_{m_\gamma \Lambda}\lrb{\bfit{r},t} &= \sum_{q\in \{x,y,z\}}  \bfit{E}_{m_\gamma \Lambda}^{q\uparrow}\lrb{\bfit{r},t}+\bfit{E}^{q\downarrow}_{m_\gamma \Lambda}\lrb{\bfit{r},t}
\end{align}
where $\bfit{E}^{q\uparrow}_{m_\gamma \Lambda}$ and $\bfit{E}^{q\downarrow}_{m_\gamma \Lambda}$, with $q\in\{x,y,z\}$, denote the scattering fields for orientations aligned parallel and anti-parallel to the $q$-axis, respectively. To modulate the relative distribution of each configuration, we vary their respective number densities in the calculation of each effective thickness.

We begin by considering an equal distribution of all configurations, i.e. the number density for each orientation is set to $N_{q\uparrow}=N_{q\downarrow}=\frac{N}{6}$. In Fig.~\ref{fig:allconfigouniformosci}, the resulting transverse intensity profile appears radially symmetric. This symmetry can be explained by the equal weighting of both parallel and anti-parallel aligned configurations, which cancels out any asymmetry in the overall intensity distribution. 
Despite this spatial symmetry, the intensity profile shows a distinct time dependence. Initially, the profile resembles that of the input pulse, characterized by lower intensity at the center and enhanced intensity at the rings ($t=0.01 \text{ ms, } t=2.5 \text{ ms, } t=5.5 \text{ ms, } t=7 \text{ ms}$ in Fig.~\ref{fig:allconfigouniformosci}). Over time, this pattern inverts. The central region becomes highly intense while the outer rings diminish ($t=1.11 \text{ ms, } t=4.15 \text{ ms}$ in Fig.~\ref{fig:allconfigouniformosci}). The system then oscillates between these two states on the time scale of the dynamical beats, which are responsible for this behaviour. This originates from the different beat frequencies associated with the individual orientations. In particular, the contribution form nuclei with EFG aligned anti-parallel to the $z$-axis produces enhanced intensity in the beam center. When the initially dominant contributions decay due to their faster dynamical oscillations, the central component becomes temporarily dominant. As the beat cycles continue, the relative weights interchange periodically, leading to a global oscillation of the transverse structure. 

Next, we examine the case of a non uniform distribution of EFG-orientations. Specifically, we choose a number density of $N_{x\uparrow}=\frac{N}{2}$ for the parallel aligned $x$-axis configuration, and $N_{q\uparrow}=N_{q\downarrow}=\frac{N}{10}$ for each of the remaining orientations. In this case, the transverse profile develops a time dependent intensity gradient, see Fig.~\ref{fig:allconfigxweightedosci}. The dominant contribution $E_{m_\gamma \Lambda}^{x\uparrow}$ initially determines the overall structure due to its larger effective thickness ($t=0.01 \text{ ms}$ in Fig.~\ref{fig:allconfigxweightedosci}). However, the increased thickness also implies a higher dynamical beat frequency. The largest contribution therefore decreases earlier in time, allowing the remaining orientations, most prominently $E_{m_\gamma \Lambda}^{x\downarrow}$, to become comparatively stronger ($t=0.43 \text{ ms}$ in Fig.~\ref{fig:allconfigxweightedosci}). Since the latter produces an opposite transverse gradient, the direction of the overall gradient reverses ($t=1.41 \text{ ms}$ in Fig.~\ref{fig:allconfigxweightedosci}). As the beat cycle proceeds, the relative dominances alternates, resulting in an oscillation of the gradient direction ($t=2.67 \text{ ms, } t=3.93 \text{ ms, and } t=6.59 \text{ ms}$ in Fig.~\ref{fig:allconfigxweightedosci}).

Thus, from this feature in an intensity profile, we would be able to deduce $N_{x\uparrow}\neq N_{x\downarrow}$. In a similar fashion, we could infer $N_{y\uparrow}\neq N_{y\downarrow}$ from these features. In this case, the gradient would result from a superposition of the $x$- and $y$-contributions, lying between the $x$- and $y$-axes. If $N_{x\uparrow}$ or $ N_{x\downarrow}$ is larger than $N_{y\uparrow}$ or $ N_{y\downarrow}$, the gradient shifts closer to the $y$-axis. Contrary, if the $y$-contributions dominate, it shifts closer to the $x$-axis. Thus, the relative distribution of quantization axes along $x$ and $y$ in the crystal can be qualitatively determined.
Using this approach, there is no visible feature that gives insights about the relative distribution of quantization axes along the $z$-axis. However, this information can still be extracted indirectly. By rotating the crystal such that the original $z$-axis aligns with the original $x$-axis, nuclei that were initially in the parallel orientation are transformed into an $x$-orthogonal orientation. In this rotated configuration, the transverse intensity profile exhibits signatures that can be used to determine the relative distribution of quantization axes along $y$- and $z$-directions. Similarly, rotating the $z$-axis into the $y$-axis allows determination of the distribution between the $x$- and $z$-aligned quantization axes. 
Hence, coherent pulse propagation of a Bessel beam offers the potential to be used as a diagnostic tool for anisotropic nuclear systems.

\subsubsection{Non paraxial regime}
 \begin{figure*}
    \centering
    \includegraphics[width = \textwidth]{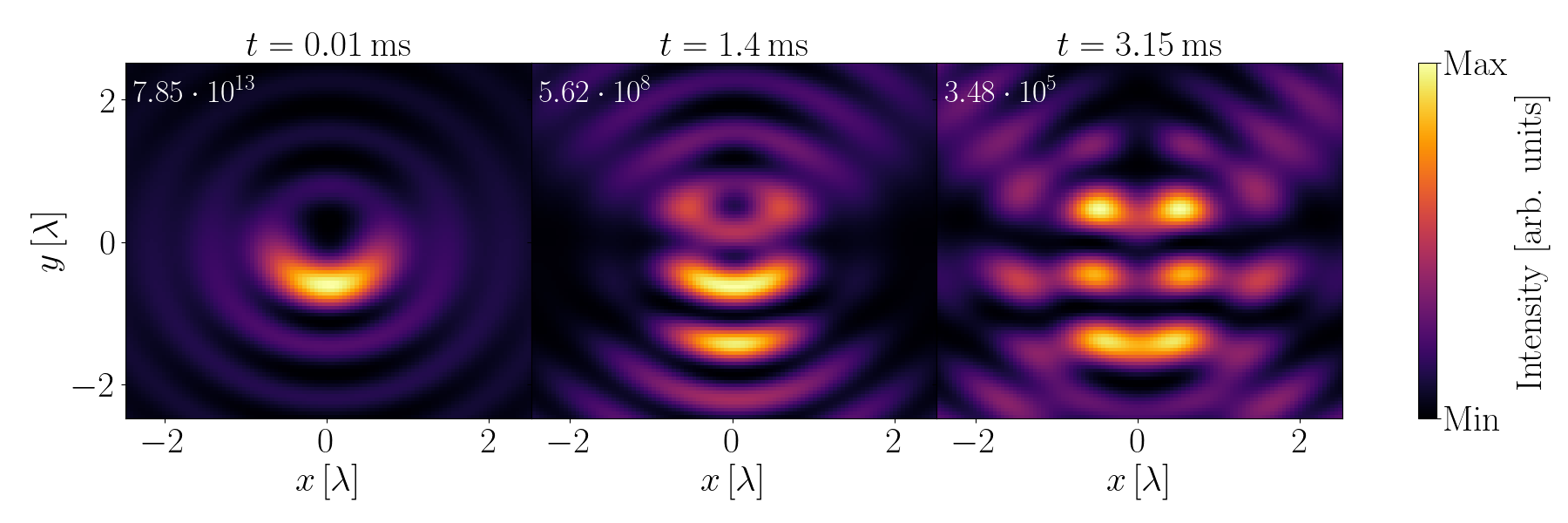}
    \caption{Transverse intensity profile for a Bessel beam propagating orthogonal to the quantization axis at different times ($\theta_k=45^\circ$). The complex structures indicate that the propagated field in the non-paraxial regime can be interpreted as a superposition of Bessel beams with different vortices. Respective maximum intensities are depicted in the upper left corner in arb. units.}
    \label{fig:nonparaxial}
\end{figure*}

From Eq.~\eqref{eq:modulusnuccurrentx} it follows that the azimuthal dependence of the nuclear currents is modulated by $\sin^2 \lrb{\theta_k}$. Thus, one expects a more pronounced spatial dependence of the dynamical beats for larger opening angles. In order to investigate this scenario, we consider the $x$-orthogonal orientation for large opening angles, i.e. $\theta_k=45^\circ$. In this regime, the transverse intensity profile exhibits stronger temporal fluctuations on the time scale of a dynamical beat, see Fig.~\ref{fig:nonparaxial}. To understand the orgin of these dynamics, it is instructive to expand the analytical solution in terms of individual scattering orders rather than working with the closed Bessel function form
\begin{align}
    &\frac{J_1\lrb{2 \sqrt{\xi \tau |\bfit{j}(\bfit{k})|^2}}}{\sqrt{\xi \tau |\bfit{j}(\bfit{k})|^2}} = \sum_{m=0}^{\infty} \frac{(-1)^m}{m! (m+1)!} \lrb{\xi \tau}^m \lrb{|\bfit{j}(\bfit{k})|^2}^m .
\end{align}
and inserting the expression for the transition from $m_g=5/2$ to $m_e=3/2$
\begin{align}
    |\bfit{j}(\bfit{k})|^2=\frac{1}{8} \lrb{2 \cos^2 \theta_k + (3+ \frac{e^{i 2 \alpha_k}+e^{-i2\alpha_k}}{2})\sin^2 \theta_k}
\end{align}
one obtains for each scattering order powers of terms containing $e^{\pm i2 \alpha_k}$. After inserting this expansion into the integral over the Bessel cone, the first scattering order ($m=0$) keeps the original vortex $m_\gamma$. In contrast, already the second scattering order ($m=1$) generates additional contributions proportional to $e^{i \lrb{m_\gamma \pm 2}\alpha_k}$. Higher scattering orders produce correspondingly higher harmonics $e^{i\lrb{m_\gamma+2m}\alpha_k}$. This structure indicates that each scattering order gives rise to additional Bessel beam components with TAM ($m_\gamma+2m$) and ($m_\gamma -2m$). The propagated field in the non-paraxial regime can therefore be interpreted as a superposition of Bessel beams with different vortices. 

Note that the same mechanism is present in the paraxial regime. There, however, the prefactor $\sin ^2\theta_k$ suppresses the azimuthal dependence, so that the higher order OAM components remain strongly supressed. They become visible only when the contribution of the scattered light with TAM $m_\gamma$ is reduced during a dynamical beat, rendering it comparable in magnitude to the components with $m_\gamma\pm2$.

\subsection{Broadband excitation}

In this Section, we want to discuss the excitation of all hyperfine split levels by considering an incoming pulse of broadband character. In contrast to the previous analysis, we must now include the summation over $l$ in Eq.~\eqref{eq:IWEplane} to calculate the scattering field of a plane wave. For this scenario, an analytical solution does not exist and we need to solve the IWE numerically for sufficiently high iteration orders. \newline
For the single-transition case, we focused on time scales on the order of $~\sim \text{ms}$, where the intensity time spectrum exhibits a dynamical beat, to investigate potential differences in the propagation behaviour between Bessel beams and plane waves. In the current analysis of a multi-level system, however, we focus on the effects of quantum beats, which arise due to the various energy differences between the possible transitions. This type of beats occur on much shorter time scales on the order of $t\sim\text{ns}$. Therefore, we will restrict our analysis to this time domain. \newline
The parameters of our crystal and Bessel beam remain equal to the single transition case, i.e. $\xi\approx6\cdot 10^7$, $\Gamma=10^{-3} \frac{1}{s}$, $\Lambda=-1$, $m_\gamma=1$ and $\theta_k=5^\circ$. It is assumed that all levels are equally populated in order to apply the same effective thickness to each transition.
\subsubsection{EFG parallel to beam propagation axis}
\begin{figure}[h]
    \centering
    \subfloat[\label{fig:zconfigbroadtransintensity}]{\includegraphics[width=\linewidth]{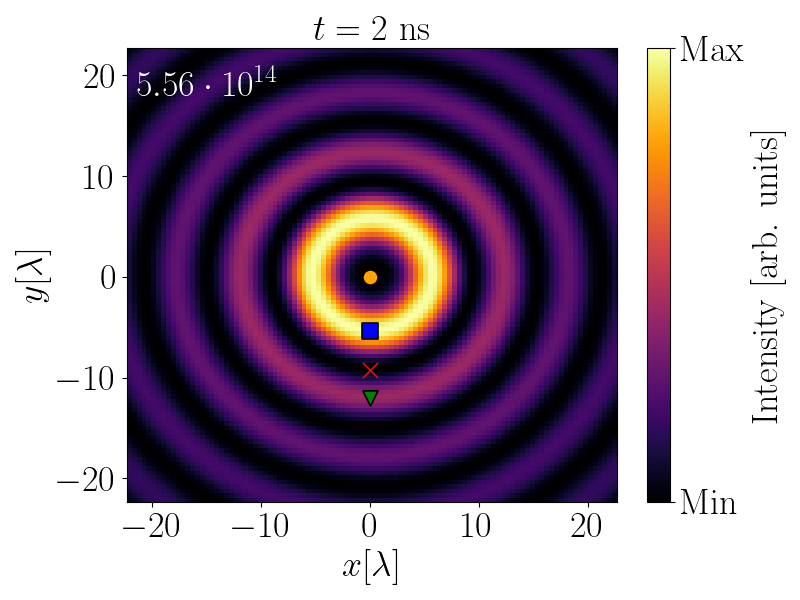}}
    \vspace{1em}
    \subfloat[\label{fig:zconfigbroadtimespectra}]{\includegraphics[width=\linewidth]{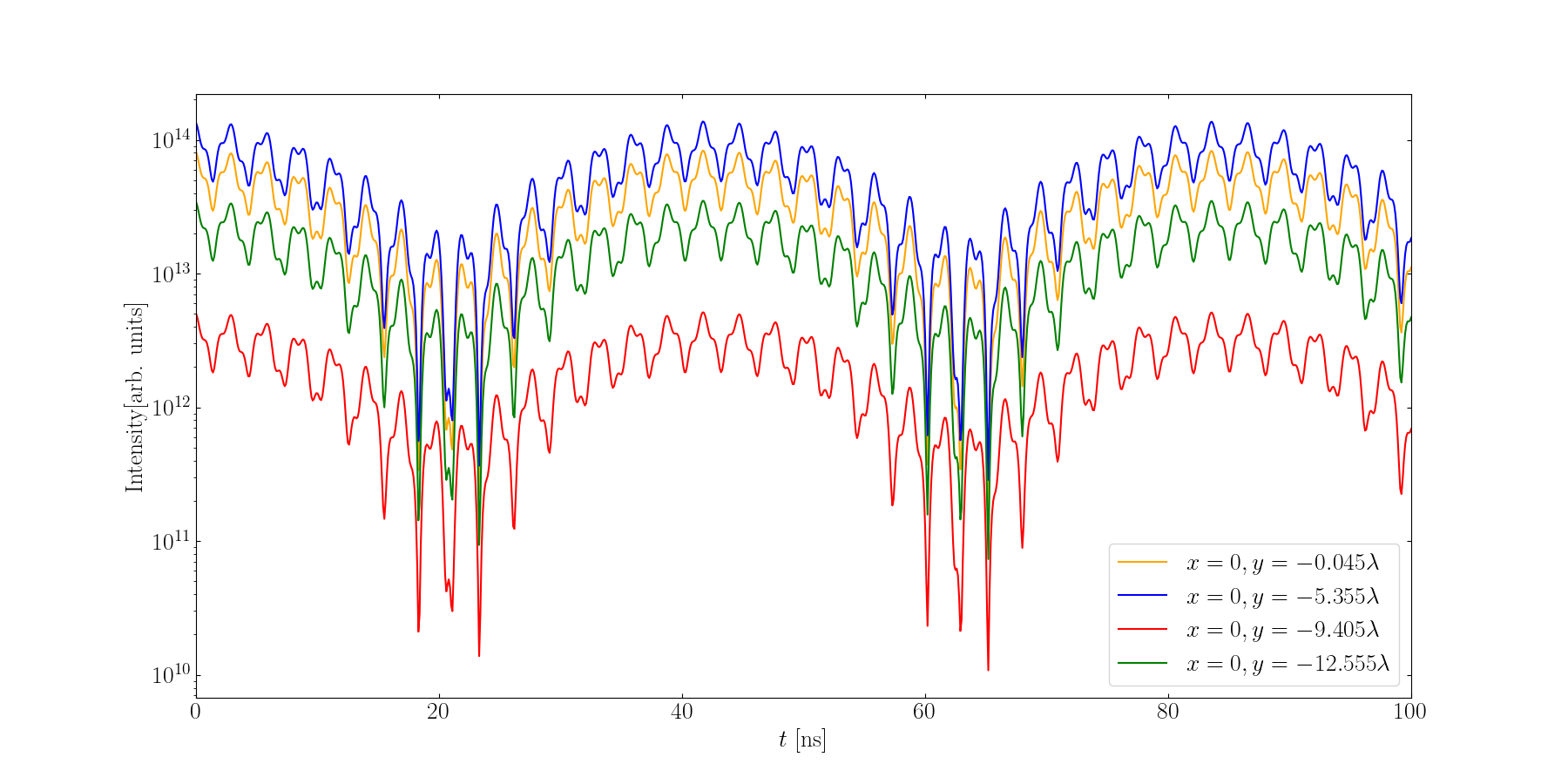}} 
    \caption{(a) shows transverse intensity profiles at different times for a Bessel beam propagating parallel to the quantization axis ($\theta_k=5^\circ$). Maximum intensity is depicted in the upper left corner in arb. units. In (b) intensity time spectra are shown at the marked positions in the transverse plane (yellow circle: $b=-0.045 \lambda$, blue square: $b=-5.355 \lambda$, red cross: $b=-9.405 \lambda$, green triangle: $b=-12.555 \lambda$, respectively. One observes no spatial dependence of the quantum beats.}
    \label{fig:zwei-bilder}
\end{figure}
\begin{figure}[t]
    \centering
    \subfloat[\label{fig:xconfigbroadtransintensity}]{\includegraphics[width=\columnwidth]{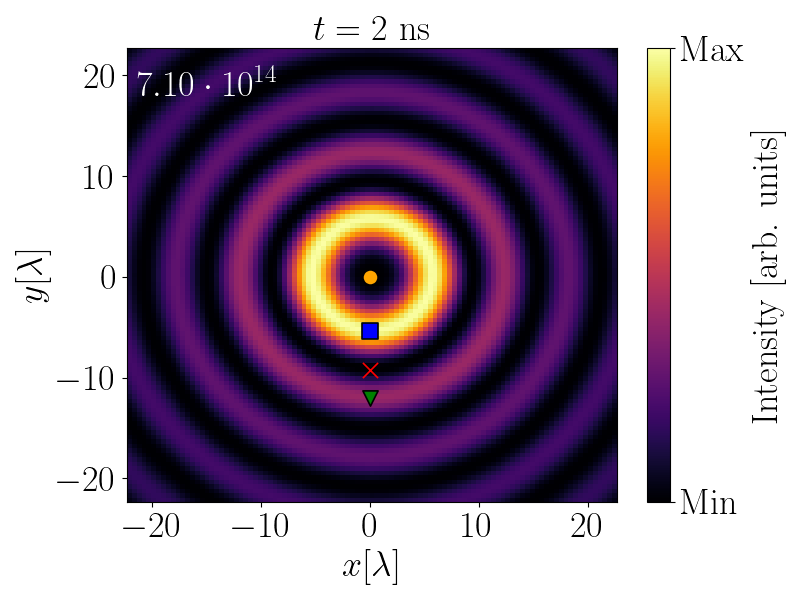}}
    \vspace{1em}
    \subfloat[\label{fig:xconfigbroadtimespectra}]{\includegraphics[width=\columnwidth]{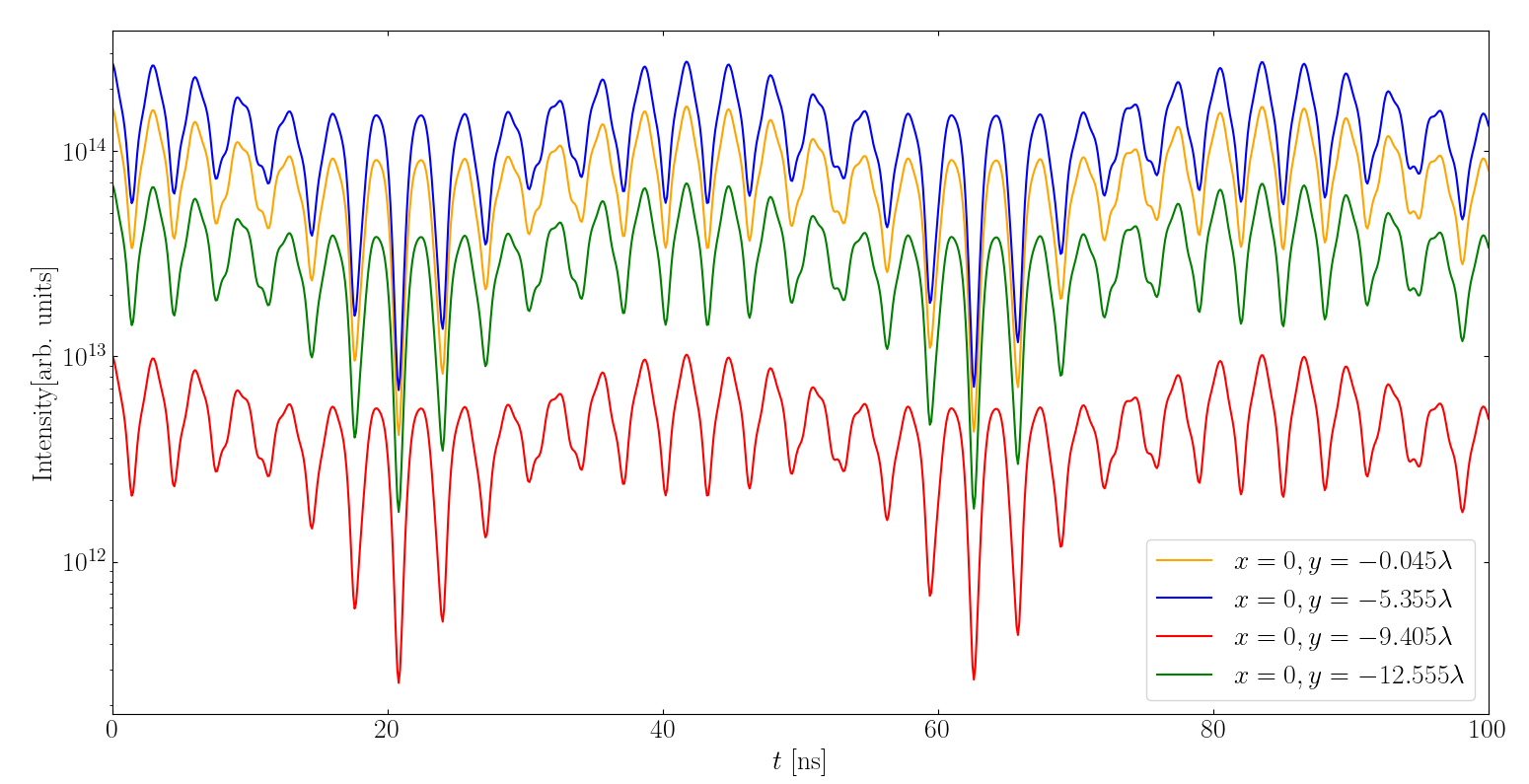}} 
    \caption{(a) shows transverse intensity profiles at different times for a Bessel beam propagating orthogonal to the quantization axis ($\theta_k=5^\circ$). Maximum intensity is depicted in the upper left corner in arb. units. In (b) intensity time spectra are shown at the marked positions in the transverse plane (yellow circle: $b=-0.045 \lambda$, blue square: $b=-5.355 \lambda$, red cross: $b=-9.405 \lambda$, green triangle: $b=-12.555 \lambda$, respectively. One observes no spatial dependence of the quantum beats.}
    \label{fig:zwei-bilder}
\end{figure}
\begin{figure*}
    \centering
    \includegraphics[width = \textwidth]{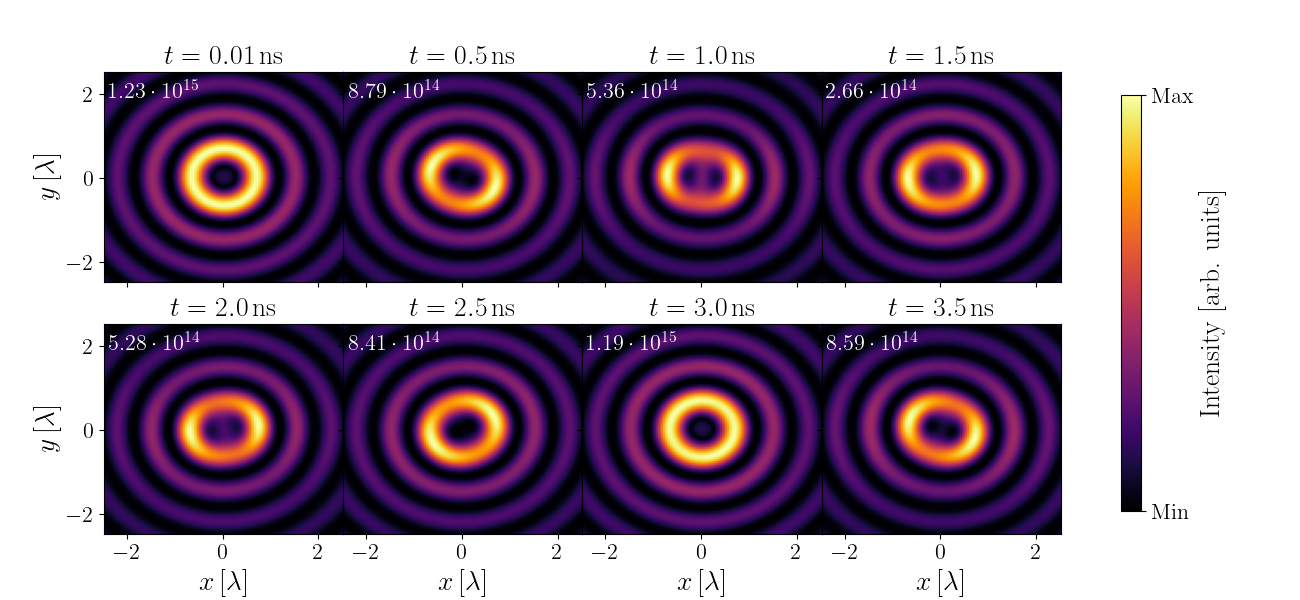}
    \caption{Transverse intensity profiles are shown at different times for a Bessel beam propagating orthogonal to the quantization axis ($\theta_k=45^\circ$). Temporal fluctuations and rotations of the intensity profile are observed. Respective maximum intensities are depicted in the upper left corner in arb. units.}
    \label{fig:nonparaxialbroad}
\end{figure*}
As in the two-level case, we analyze the possible EFG orientations individually, beginning with the configuration where the quantization axis is aligned with the $z$-direction. The resulting transverse intensity distribution is shown in in Fig.~\ref{fig:zconfigbroadtransintensity}. The profile closely resembles the profile we obtained for the single $\Delta m=-1$ transition. Although all allowed transitions ($\Delta m=0,\pm1$) are in principle driven by the broadband excitation, the $\Delta m=0$ and $\Delta m=+1$ channels are strongly suppressed in the paraxial regime as can be deduced from the expressions for the interaction matrix hamiltonian in Appendix~\ref{app:interaction}. As a consequence, their contribution to the transverse intensity profile is negligible, and the spatial structure is dominated by the $\Delta m=-1$ transitions. The time spectrum exhibits clear quantum beats originating from the interference of the different hyperfine transition frequencies. The time spectra evaluated at different radial coordinates coincide up to an overall amplitude factor, demonstrating that the spatial profile does not change in time.
\subsubsection{EFG perpendicular to beam propagation axis}
We now proceed to the case of orthogonal orientations between the propagation direction and the quantization axis. As in previous sections, we restrict ourselves to the configuration where the quantization axis points in $x$-direction. The resulting transverse intensity profile is shown in Fig.~\ref{fig:xconfigbroadtransintensity}. Remarkably, the spatial structure is indistinguishable from that obtained in the parallel configuration. It arises from the superposition of all allowed transitions. In this case, the $\Delta m=0,\pm1$ transitions contribute with comparable strength. Individually, $\Delta m=\pm1$ produce an asymmetric transverse profile in the orthogonal orientation. However, their superposition restores the radial symmetry.
Despite the spatial similarity to the parallel case, the time spectra differ from those of the parallel configuration. The quantum beat structure reflects a different set of driven transitions and their respective interference frequencies. Hence, the equality of the spatial profiles does not imply identical excitation dynamics.

Since both parallel and orthogonal orientations yield radially symmetric intensity profiles under broadband excitation in the paraxial regime, their superposition does not generate additional spatial signatures. Consequently, unlike in the single transition scenario, a broadband Bessel beam excitation does not provide a diagnostic tool for distinguishing $180^\circ$ doping configurations via the transverse intensity profile.

\subsubsection{Non paraxial regime}
As in the single transition case, the non paraxial regime introduces additional features to the scattering dynamics. To illustrate these effects, we consider the $x$-orthogonal configuration.
Fig.~\ref{fig:nonparaxialbroad} shows the corresponding transverse intensity profile at different times. In contrast to the paraxial regime, one observes fluctuations between an annular and an asymmetric profile. In addition, the asymmetric profile appears to rotate in the transverse plane. The fluctuation arise because, in the non paraxial regime, the superposition of only the $\Delta m=\pm1$ contributions does not yield a symmetric intensity distribution. The formation of an annular pattern requires the additional contribution from the $\Delta m=0$ transitions. Thus, the fluctuations reflect the changing relative intensity of the different transition channels.
The rotation of the asymmetric profile originates from phase differences between transitions with the same $\Delta m$. Although these produce qualitatively similar spatial profiles, their relative phase factors lead to a rotation of the resulting intensity pattern in the transverse plane.

\section{Conclusion}\label{sec:fin}
We have developed a generalized formulation of the IWE for the propagation of Bessel beams in resonant nuclear media. By expressing the structured field as a coherent superposition of circularly polarized plane waves on a conical momentum distribution, the IWE can be applied to each plane-wave component individually while accounting for the modified effective thickness due to the propagation geometry. This approach provides a systematic framework for describing multiple scattering of twisted light in nuclear hyperfine-split systems.

For a single transition, the spatial and temporal dynamics strongly depend on the relative orientation between the beam propagation direction and the quantization axis. If both axes are aligned, the azimuthal integration over the Bessel cone preserves rotational symmetry and yields spatially homogeneous dynamical beats across the transverse plane. In contrast, for orthogonal orientations the coupling strength acquires an explicit azimuthal dependence, leading to position-dependent beat frequencies and spatially modulated intensity patterns. In the non paraxial regime, multiple scattering with orthogonal orientation generates beams with higher order OAM components, revealing a non trivial intensity profile resulting from their superposition.

When all hyperfine transitions are excited simultaneously by a broadband pulse, the situation changes qualitatively. In the paraxial regime, the coherent superposition of the $\Delta m=0,\pm1$ channels restores radial symmetry even for orthogonal orientations. Although the time spectra exhibit quantum beats arising from the hyperfine energy splittings, the transverse intensity profile becomes largely insensitive to the orientation of the quantization axis. Consequently, the diagnostic signatures that are present in the single transition scenario are suppressed under broadband excitation. However, in the non paraxial regime, the orthogonal orientation only preserves the radial symmetry if all channels contribute equally which leads to temporal fluctuations of the intensity pattern. 

Our results demonstrate that the interplay between twisted light and coherent pulse propagation leads to qualitatively different propagation regimes depending on the spectral selectivity of the excitation. In particular, coherent pulse propagation of a Bessel beam can generate additional vortices and encode microscopic orientation information into the time and space dependent intensity pattern. These effects are absent for plane wave excitation and highlight the potential of vortex light beams for probing anisotropic nuclear systems.


\begin{acknowledgments}

This research was supported by the Austrian Science Fund (FWF) [grant DOI:10.55776/F1004] (COMB.AT) together with the German Science Foundation (Deutsche Forschungsgemeinschaft, DFG), Project No.~534051539.
A.P. gratefully acknowledges the Heisenberg
Program of the DFG (Project No.~435041839).

\end{acknowledgments}

\appendix

\section{Nucleus-Bessel beam interaction}
\label{app:interaction}

Here, we provide a brief theoretical description of the interaction between a Bessel beam and a nucleus. The interaction Hamiltonian is given by 
\begin{align}
    \hat{H}^{B}_{int}(\bfit{b})=-\frac{1}{c} \int \bfit{j}\lrb{\bfit{r}} \bfit{A}_{\zeta m_\gamma}\lrb{\bfit{r}-\bfit{b},t} d^3\bfit{r} \ .
\end{align}
Note that we need to introduce the impact parameter 
\begin{align}
    \bfit{b}= \begin{pmatrix}
        b \cos(\alpha_b) \\
        b  \sin(\alpha_b) \\
        0
    \end{pmatrix} \,
\end{align}
 to take into account the spatially inhomogeneous profile of a Bessel beam. Making use of the momentum representation of the Bessel beam in Eq.~\eqref{besselbeam}, the interaction Hamiltonian can be written as 
\begin{align}
    &\hat{H}^{B}_{int}(\bfit{b})=\frac{A_0}{\sqrt{2\pi}} e^{-i\omega t}  \sum_{L=1}^\infty \sum_{M=-L}^L \sqrt{2L+1}i^{L-m_\gamma} \notag \\ 
    &\times \int \bfit{j}\lrb{\bfit{r}} \lrb{\bfit{\mathcal{A}}^\mathcal{M}_{LM}(\bfit{r})+i\Lambda\bfit{\mathcal{A}}^\mathcal{E}_{LM}(\bfit{r})} d^3\bfit{r} \notag \\
    &\times  \int_0^{2\pi} D_{M,\Lambda}^L(\alpha_k,\theta_k,0) e^{i m_\gamma\alpha_k} e^{-i \zeta b \cos\lrb{\alpha_k - \alpha_b}} \lrb{\bfit{k_\perp}}d\alpha_k \ ,
\end{align}
where $D^L_{M,\Lambda}\lrb{\alpha_k,\theta_k,0}$ is the Wigner-$D$-function and $\bfit{\mathcal{A}}^\mathcal{M}_{LM}$ and $\bfit{\mathcal{A}}^\mathcal{E}_{LM}$ are the magnetic and electric multipole fields of the multipole expansion of a plane wave, respectively.
We factorize $D_{M,\Lambda}^L(\alpha_k,\theta_k,0)=e^{-i M \alpha_k}d^L_{M,\Lambda}\lrb{\theta_k}$ and use the representation of the Bessel function in Eq.~\eqref{besselrep} to simplify
\begin{align}
    \hat{H}^{B}_{int}(\bfit{b})&=\frac{A_0}{\sqrt{2\pi}} e^{-i\omega t}  \sum_{L=1}^\infty \sum_{M=-L}^L \sqrt{2L+1}i^{L+M-2m_\gamma} \notag \\ 
    &\times \int \bfit{j}\lrb{\bfit{r}} \lrb{\bfit{\mathcal{A}}^\mathcal{M}_{LM}(\bfit{r})+i\Lambda\bfit{\mathcal{A}}^\mathcal{E}_{LM}(\bfit{r})} d^3\bfit{r} \notag \\
     &\times d^L_{M,\Lambda}\lrb{\theta_k} e^{i (m_\gamma - M)\phi_b} J_{m_\gamma-M}\lrb{\zeta b} \ .
\end{align}
With the aid of \cite{AP_dipoleforbidden,ring2004nuclear}, we can evaluate the integral
\begin{align}
    M_{LM}^{\mu}= \int \bfit{j}\lrb{\bfit{r}} \bfit{\mathcal{A}}^\mathcal{\mu}_{LM}(\bfit{r}),
\end{align}
where $M_{LM}^{\mu}$ is the multipole moment of type $\mu\in{\mathcal{M},\mathcal{E}}$.
Typically, only one or two multipole orders dominate in the sum over $L$. In this work, we consider a magnetic dipole character for the nuclear transitions. Thus, we consider only the magnetic terms in the derivation of the interaction matrix element between two nuclear states $\mathcal{V}_{eg}:=\bra{I_em_e}\hat{H}_{int}\ket{I_g m_g}_r$. 

Since we consider multiple orientations between propagation direction and quantization axis in this work, we need to express the nuclear states $\ket{I_lm_l}_r$ defined with respect to a quantization axis, pointing in an arbitrary direction, in terms of the nuclear states in the parallel configuration $\ket{I_lm_l}_p$ using Wigner-$D$-matrices \cite{edmonds1996angular}
\begin{align}
    \ket{I_l m_l}_r = \sum_{m_l'} D^{I_l}_{m_l,m_l'}\lrb{\alpha,\beta,\gamma} \ket{I_l m_l'}_p \ ,
\end{align}
where $\alpha,\beta$ and $\gamma$ are the Euler angles describing the rotation of the propagation axis to the quantization axis. Applying this transformation to the nuclear states the interaction matrix element reads
\begin{align}
    \mathcal{V}_{eg}=\sum_{m_e',m_g'} D^{I_e}_{m_e,m_e'}\lrb{\alpha,\beta,\gamma}^* D^{I_g}_{m_g,m_g'}\lrb{\alpha,\beta,\gamma} \notag \\ \cross \bra{I_e m_e'}\hat{H}_{int}\ket{I_g m_g'}_p .
\end{align}
Applying the Wigner-Eckart theorem \cite{ring2004nuclear}, the matrix element factorises as
\begin{align}
    \bra{I_em_e'} M_{LM}^\mathcal{M} \ket{I_g m_g'}=\frac{1}{\sqrt{2L+1}} \braket{I_e m_e' \ I_g-m_g'}{L M} \notag \\ \cross \bra{I_e \mid } M_{L}^\mathcal{M} \ket{\mid I_g} \ ,
\end{align}
where $\braket{I_e m_e' \ I_g-m_g'}{L M}$ are Clebsch-Gordan coefficients and the reduced matrix element $\bra{I_e \mid } M_{L}^\mathcal{M} \ket{\mid I_g}$ is related to the magnetic reduced transition probability $\mathcal{B}\lrb{\mathcal{M}L, I_g\rightarrow I_e}=\frac{1}{2I_g+1}\abs{ \bra{I_e \mid } M_{L}^\mathcal{M} \ket{\mid I_g}}^2$.  

Collecting terms and using the following identity for Wigner-$D$ matrices \cite{varshalovich1988quantum}
\begin{align}
    \sum_{m_e',m_g'} &D^{I_e}_{m_e,m_e'}\lrb{\alpha,\beta,\gamma}^* D^{I_g}_{m_g,m_g'}\lrb{\alpha,\beta,\gamma} \braket{I_e m_e' \ I_g-m_g'}{1 M} \notag \\ = &\braket{I_e m_e \ I_g-m_g}{1 m_e-m_g} D^{1}_{m,m_e-m_g}\lrb{\alpha,\beta,\gamma} \ ,
\end{align} 
we arrive at the final expression for the magnetic multipole transition matrix element 
\begin{align}
    \mathcal{V}_{eg} \lrb{\bfit{b}}&= -\sqrt{\frac{8 \pi \mathcal{I}}{9 c \epsilon_0}} \sqrt{2I_g+1} \sqrt{\mathcal{B}\lrb{\mathcal{M}1, I_g\rightarrow I_e}} \notag \\
    &\cross \lrb{-1}^{I_g-m_g}  \braket{I_e m_e \ I_g-m_g}{1 m_e-m_g}  \notag \\
    &\cross \sum_M \lrb{-i}^{2m_\gamma-M} e^{i\lrb{m_\gamma-M}\alpha_b} \notag \\
    &\cross J_{m_\gamma-M}\lrb{\zeta b} d_{M,\Lambda}^1\lrb{\theta_k } D^{1}_{M,m_e-m_g}\lrb{\alpha,\beta,\gamma} \ , \label{eq:besselintnonparallel} 
\end{align}

The $^{229}$Th clock transition proceeds mainly via the $\mathcal{M}1$-channel with a negligible $\mathcal{E}2$ multipole mixing \cite{Kirschbaum2022} which is be disregarded for the scope of this work. Hence, in the comparisons with the scattered intensities which are obtained using the IWE, we solely consider the matrix element for a magnetic dipole transition.

\section{Decoherence effects}
\label{app:decoherence}

In the crystal environment, the thorium nucleus experiences magnetic dipole interactions with randomly oriented spins of the surrounding nuclei. These interactions lead to additional loss in the form of decoherence rates  $\gamma_{ij}^c$ between two levels $i$ and $j$ \cite{kazakov2012performance}. In order to incorporate this effect in the framework of the IWE, we make use of the alternative formalism of the Maxwell Bloch equations (MBE). This approach describes the problem by using a set of partial differential equations derived from the optical Bloch and Maxwell's equations \cite{scully1999quantum}. For a simple two-level system they read
\begin{align}
    \frac{d \hat{\rho}}{d t} &= -\frac{i}{\hbar}\lrsb{\hat{H},\hat{\rho}} + \mathcal{L}, \\
    \lrb{\frac{\partial}{\partial z} + \frac{1}{c} \frac{\partial}{\partial t}} \Omega\lrb{z,t} &= i \frac{2 \xi \Gamma}{L} \rho_{21}\lrb{z,t}  ,
\end{align}
where $\hat{H}$ and $\hat{\rho}$ denote the Hamiltonian and density matrix of the two-level system, respectively. The field intensity is given by the modulus squared of the Rabi frequency $\Omega$. Here, the decoherence rates can be included straightforward in the Lindblad term $\mathcal{L}$, by adding terms of the form $\mathcal{L}^c_{ij}=(1-\delta_{ij})\gamma^c_{ij} \rho_{ij}$ \cite{scully1999quantum,shore1991theory}. 

Assuming a weak light interaction, one can use a perturbative ansatz to solve the MBE. With the same initial conditions as we used in the IWE case, the solution is given by  \cite{liao2014all}
\begin{align}
    \Omega_p\lrb{z,t} = e^{-\frac{\Gamma + 2\gamma_{ij}^c}{2} t} \delta(t) - \frac{\xi \Gamma z}{L} \frac{J_1\lrb{2\sqrt{\frac{\xi \Gamma t z}{L}}}}{ \sqrt{\frac{\xi \Gamma t z}{L}}} e^{-\frac{\Gamma + 2\gamma_{ij}^c}{2} t}
\end{align}
A comparison with the IWE solution reveals that for the two level system the decoherence affects only the decay rate. Unfortunately, for a multi level excitation this is not true anymore. But since we discuss the multi level excitations on time scales $\sim$ ns and the decoherence rates are in the order of $100$ Hz, we can neglect them in these cases. 

The decoherence rates for each allowed magnetic dipole transition in the quadrupole-split level scheme are given by \cite{kazakov2012performance}
\begin{align}
    &\gamma^c_{\frac{5}{2},\frac{3}{2}}=\gamma^c_{-\frac{5}{2},-\frac{3}{2}} \approx 2\pi \cdot 251 \text{ Hz}, \\
    &\gamma^c_{\frac{3}{2},\frac{3}{2}}=\gamma^c_{-\frac{3}{2},-\frac{3}{2}} \approx 2\pi \cdot 108 \text{ Hz},\\
    &\gamma^c_{\frac{1}{2},\frac{3}{2}}=\gamma^c_{-\frac{1}{2},-\frac{3}{2}} \approx 2\pi \cdot 158 \text{ Hz},\\
    &\gamma^c_{\frac{3}{2},\frac{1}{2}}=\gamma^c_{-\frac{3}{2},-\frac{1}{2}} \approx 2\pi \cdot 84 \text{ Hz},\\
    &\gamma^c_{\frac{1}{2},\frac{1}{2}}=\gamma^c_{-\frac{1}{2},-\frac{1}{2}} \approx 2\pi \cdot 150 \text{ Hz},\\
    &\gamma^c_{-\frac{1}{2},\frac{1}{2}}=\gamma^c_{\frac{1}{2},-\frac{1}{2}} \approx 2\pi \cdot 142 \text{ Hz}.
\end{align}

\section{Single transition with $\Delta m=0,1$} \label{app:deltam0}
\begin{figure}[h]
    \centering
    \subfloat[\label{fig:deltapluscomp}]{\includegraphics[width=\linewidth]{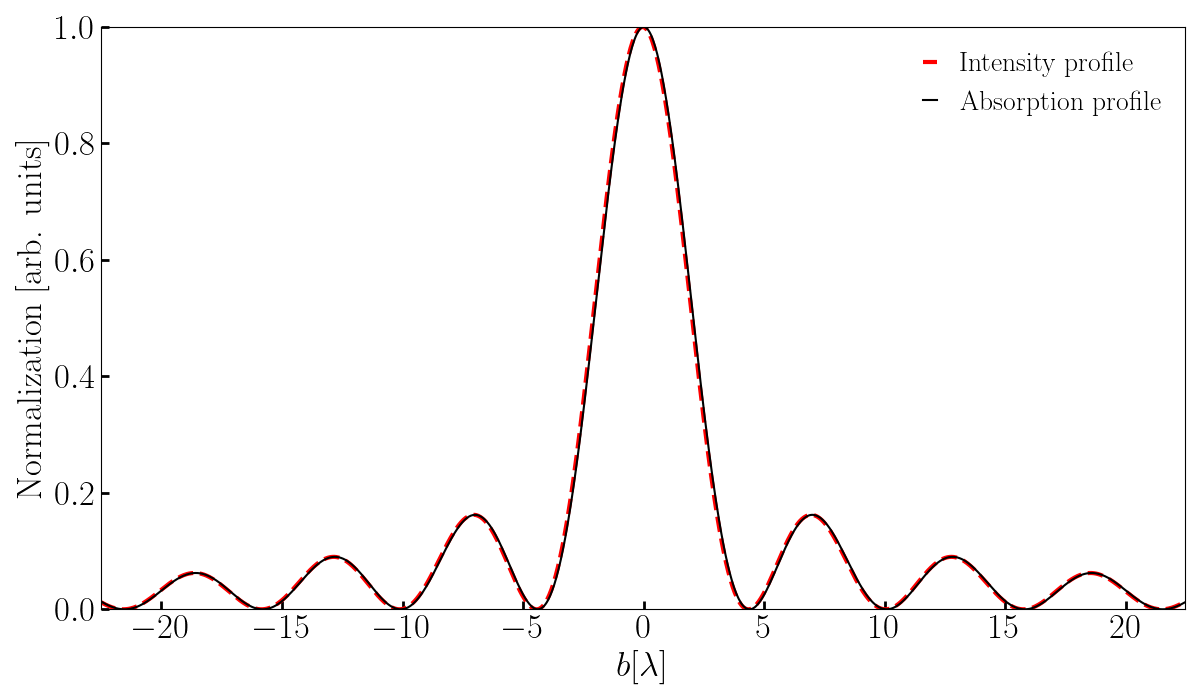}}
    \vspace{1em}
    \subfloat[\label{fig:delta0comp}]{\includegraphics[width=\linewidth]{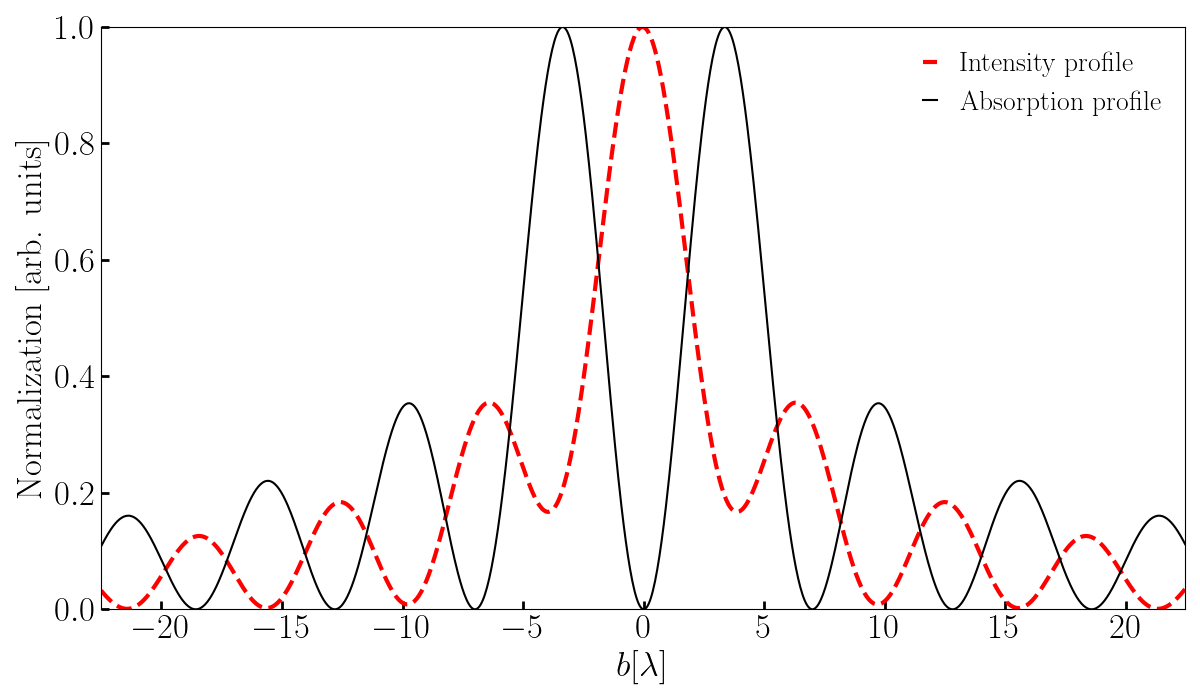}} 
    \caption{The normalized intensity profile is compared to the normalized absorption profile for a (a) $\Delta m=1$- and (b) $\Delta m=0$-transition. The intensity profile does not change in time on a qualitative level and the beam and crystal parameters are the same as in Sec.~\ref{sec:singlepar}.}
    \label{fig:zwei-bilder}
\end{figure}

In Sec.~\ref{sec:singlepar}, we saw that the result of the IWE has the same spatial dependence as the interaction between a nucleus and a Bessel beam. This holds also for a transition with $\Delta m=1$, e.g. from $\ket{\frac{5}{2},\frac{1}{2}}$ to $\ket{\frac{3}{2},\frac{3}{2}}$, which is illustrated in Fig.~\ref{fig:deltapluscomp}.
However, the correspondence breaks down in the case of $\Delta m=0$. In order to see this, we choose the transition from $\ket{\frac{5}{2},\frac{3}{2}}$ to $\ket{\frac{3}{2},\frac{3}{2}}$. A comparison of the scattered intensity and the absorption profile are shown in Fig.~\ref{fig:delta0comp}. The curves show no correspondence. We assume that the origin of this discrepancy lies in the emission process. The interaction matrix profile describes only the absorption of a Bessel beam. Thus, it is more a feature of the $\Delta m=\pm1$ cases that the intensity profile is not altered by the emission.

\section{Nuclear current densities}
\label{app:currents}
In the calculation and the interpretation of the Bessel beam propagation in the case of a single transition the modulus squared of the nuclear current densities $|\bfit{j}_{l}(\bfit{k})|^2$ play a crucial role. Thus, we state here the expressions of the latter for all magnetic dipole allowed transitions in the parallel and $x$-orthogonal orientations

Parallel configuration:
\begin{equation}
\begin{aligned}
    &|\bfit{j}_{\frac{5}{2},\frac{3}{2}}(\bfit{k})|^2=|\bfit{j}_{-\frac{5}{2},-\frac{3}{2}}(\bfit{k})|^2
    =\frac{1}{8} \lrb{3+ \cos \lrb{2 \theta_k}},\\
    &|\bfit{j}_{\frac{3}{2},\frac{3}{2}}(\bfit{k})|^2=|\bfit{j}_{-\frac{3}{2},-\frac{3}{2}}(\bfit{k})|^2
    =\frac{\sin^2 \lrb{\theta_k}}{5},\\
    &|\bfit{j}_{\frac{1}{2},\frac{3}{2}}(\bfit{k})|^2=|\bfit{j}_{-\frac{1}{2},-\frac{3}{2}}(\bfit{k})|^2
    =\frac{1}{80} \lrb{3+ \cos \lrb{2 \theta_k}},\\
    &|\bfit{j}_{\frac{3}{2},\frac{1}{2}}(\bfit{k})|^2=|\bfit{j}_{-\frac{3}{2},-\frac{1}{2}}(\bfit{k})|^2
    =\frac{3}{40} \lrb{3+ \cos \lrb{2 \theta_k}},\\
    &|\bfit{j}_{\frac{1}{2},\frac{1}{2}}(\bfit{k})|^2=|\bfit{j}_{-\frac{1}{2},-\frac{1}{2}}(\bfit{k})|^2
    =\frac{3 \sin^2 \lrb{\theta_k}}{10},\\
    &|\bfit{j}_{-\frac{1}{2},\frac{1}{2}}(\bfit{k})|^2=|\bfit{j}_{\frac{1}{2},-\frac{1}{2}}(\bfit{k})|^2
    =\frac{3}{80} \lrb{3+ \cos \lrb{2 \theta_k}}.
\end{aligned}
\end{equation}
No dependence on the azimuthal angle is observed for any of the transitions. This implies no spatial dependence of the dynamical beat, and the intensity profile does not exhibit qualitative changes in time for each individual transition. 

$x$-orthogonal configuration 
\begin{equation}
\begin{aligned}
    &|\bfit{j}_{\frac{5}{2},\frac{3}{2}}(\bfit{k})|^2=|\bfit{j}_{-\frac{5}{2},-\frac{3}{2}}(\bfit{k})|^2\\
    &=\frac{1}{8} \lrb{2 \cos^2\lrb{\theta_k} + \lrb{3+\cos \lrb{2 \alpha_k}} \sin^2\lrb{\theta_k}},\\
    &|\bfit{j}_{\frac{3}{2},\frac{3}{2}}(\bfit{k})|^2=|\bfit{j}_{-\frac{3}{2},-\frac{3}{2}}(\bfit{k})|^2\\
    &=\frac {1} {5}\lrb{\cos^2 \lrb{\theta_k}+ \sin^2 (\theta_k)\sin^2 \lrb{\alpha_k}},\\
    &|\bfit{j}_{\frac{1}{2},\frac{3}{2}}(\bfit{k})|^2=|\bfit{j}_{-\frac{1}{2},-\frac{3}{2}}(\bfit{k})|^2\\
    &=\frac{1}{80} \lrb{2 \cos^2\lrb{\theta_k} + \lrb{3+\cos \lrb{2 \alpha_k}} \sin^2\lrb{\theta_k}},\\
    &|\bfit{j}_{\frac{3}{2},\frac{1}{2}}(\bfit{k})|^2=|\bfit{j}_{-\frac{3}{2},-\frac{1}{2}}(\bfit{k})|^2\\
    &=\frac{3}{20} \lrb{ \cos^2\lrb{\theta_k} +\frac{1}{2} \lrb{3+\cos \lrb{2 \alpha_k}} \sin^2\lrb{\theta_k}},\\
    &|\bfit{j}_{\frac{1}{2},\frac{1}{2}}(\bfit{k})|^2=|\bfit{j}_{-\frac{1}{2},-\frac{1}{2}}(\bfit{k})|^2\\
    &=\frac {3} {10}\lrb{\cos^2 \lrb{\theta_k}+ \sin^2 (\theta_k)\sin^2 \lrb{\alpha_k}},\\
    &|\bfit{j}_{-\frac{1}{2},\frac{1}{2}}(\bfit{k})|^2=|\bfit{j}_{\frac{1}{2},-\frac{1}{2}}(\bfit{k})|^2\\
    &=\frac{3}{40} \lrb{ \cos^2\lrb{\theta_k} +\frac{1}{2} \lrb{3+\cos \lrb{2 \alpha_k}} \sin^2\lrb{\theta_k}}.
\end{aligned}
\end{equation}
All expressions depend on the azimuthal angle. Consequently, the dynamical beat can vary spatially, which leads to temporal fluctuations of the intensity profile for each allowed transition.

\bibliography{refs}

\end{document}